\documentclass[twocolumn, superscriptaddress, prb, preprintnumbers, showpacs, floatfix, amsmath,amssymb]{revtex4-1}

\usepackage{graphicx}
\usepackage{hyperref}

\usepackage{hyperref}
\hypersetup{backref,colorlinks=true,allcolors=blue}

\usepackage{float}
\usepackage{multirow}
\usepackage{tabularx}
\usepackage[table]{xcolor}

\usepackage{tikz}
\usetikzlibrary{shapes,arrows,calc}
\usepackage{makecell}
\usepackage{cancel}
\usepackage{tcolorbox}

\usepackage[caption=false]{subfig}

\newcommand{\splitbox}[1]{%
  \linespread{1}\selectfont
  \renewcommand*{\arraystretch}{1.0}%
  \begin{tabular}{@{}l@{}l@{}}
    \strut#1\strut\strut
  \end{tabular}%
}

\def\u{DFT+\textit{U} }
\def\uu{TDDFT+\textit{U} }
\def\td#1{$_\mathrm{#1}$}
\def\tu#1{$^\mathrm{#1}$}
\def\tdu#1#2{$_\mathrm{#1}^\mathrm{#2}$}

\definecolor{lblue}{RGB}{135,206,235}
\definecolor{pink}{RGB}{233,150,122}
\definecolor{lpurple}{RGB}{221,160,221}

\def\hlp#1{\bfseries\colorbox{pink}{#1}}
\def\hlb#1{\bfseries\colorbox{lblue}{#1}}
\def\hlv#1{\bfseries\colorbox{lpurple}{#1}}

\begin{document}

\title{TDDFT+\textit{U}: Hubbard corrected approximate density-functional theory in the excited-state regime}

\author{Okan K. Orhan}
\author{David D. O'Regan}%
\affiliation{%
 School of Physics, Trinity College Dublin, Dublin 2, Ireland}

\date{\today}

\begin{abstract}
We develop a generalization of the Kohn-Sham density functional theory (KS-DFT) + Hubbard $U$ (DFT+$U$)
method to the excited-state regime. This has the form of Hubbard $U$ corrected linear-response time-dependent DFT,  or `TDDFT+$U$'.
Combined with calculated linear-response Hubbard $U$ parameters, it may provide a computationally light, first-principles method for the simulation of tightly-bound excitons on transition-metal ions.
Our presented implementation combines linear-scaling DFT+$U$ and linear-scaling TDDFT, but the approach
is broadly applicable.
In detailed benchmark tests on two Ni-centred diamagnetic coordination complexes with variable $U$ values, it is shown that the Hubbard $U$ correction to an approximate 
adiabatic semi-local 
exchange-correlation interaction kernel lowers the excitation energies of transitions exclusively within the targeted localised subspace, by increasing the exciton binding of the
corresponding electron-hole pairs. 
This partially counteracts  the Hubbard $U$ correction to the exchange-correlation potential in KS-DFT,
which increases excitation energies  into, out of,  and within the targeted localised subspace by modifying the underlying KS-DFT eigenspectrum.
This compensating effect is  most pronounced for optically dark transitions between localized orbitals 
of the same angular momentum,  for which experimental observation may be challenging and  theoretical approaches are at their most necessary.
Interestingly, we find that first-principles TDDFT+$U$ 
seems to offer a remarkably good 
agreement with experiment for a 
perfectly closed-shell complex on which approximate
TDDFT under-performs, but only when TDDFT+$U$ 
is applied to the DFT eigenspectrum
and not to the DFT+$U$ one.
In tests on an open-shell, non-centrosymmetric, high-spin cobalt coordination complex, we find that 
first-principles TDDFT+$U$ again compensates 
for the DFT+$U$ blue-shift
in $3d\rightarrow 3d$ transitions, 
but that using the DFT eigenspectrum  is not viable
due to the emergence of a singlet instability.
%
Overall, our results point to shortcomings in the 
contemporary DFT+$U$ corrective potential, 
either in its functional form, or when 
applied to transition-metal orbitals but not to ligand
ones, or both.

\end{abstract}

\maketitle
\section{Introduction}

Density-functional theory (DFT)~\cite{PhysRev.136.B864,PhysRev.140.A1133} 
provides a  computationally tractable means by which  
to investigate the quantum-mechanically derived properties of 
 molecules and materials.  
TDDFT~\cite{PhysRevLett.52.997} is its elegant extension 
to the dynamical,  excited-state regime. 
TDDFT is now widely used to investigate 
the excitation spectra of extended solids and molecules alike~\cite{PhysRevLett.76.1212,BAUERNSCHMITT1997573,doi:10.1063/1.477483}, due to its relatively low computational 
cost relative to wave-function and Green's function based approaches.
While DFT and TDDFT are both exact in principle, 
their accuracies in practice are limited 
by the  approximations  currently available
for the exchange-correlation (xc) contribution to the total-energy
functional $E_\textit{xc}$ and its derived
 interaction kernel (by second functional derivatives),  $f_\textit{xc}$.
Common xc-functionals include local functionals such 
as the local density approximation~\cite{PhysRev.140.A1133},
semi-local functionals such as  generalized gradient approximations~\cite{PhysRevLett.77.3865}, and semi-empirical functionals 
such as hybrids~\cite{doi:10.1021/j100091a024,doi:10.1063/1.478522,doi:10.1063/1.1564060}. 
In practice, an adiabatic, i.e., time-averaged interaction approximation
is made to construct the  xc-kernels 
of contemporary applied TDDFT. 
The latter is often also restricted, for  
expediency, to the 
linear-response regime appropriate only to low-energy, low-oscillator-strength excitations.

\subsection{Self-interaction error in approximate DFT
and its correction by Hubbard $U$ based methods}

Perhaps the most transparent systematic error exhibited by approximate functionals is the single-particle self-interaction error (SIE)~\cite{Cohen792}, 
i.e. the tendency of electrons to effectively self-repel, 
and has been demonstrated clearly in  single-electron systems such as the  molecule H$_2^+$~\cite{doi:10.1063/1.463297,Savin_1996,PhysRevB.56.16021,doi:10.1063/1.476859}.
This error becomes more complicated in the many-body case and hence, by necessity, there has emerged the more general concept of  many-body self-interaction error~\cite{PhysRevB.23.5048}, also known as delocalisation error~\cite{doi:10.1063/1.2403848,doi:10.1063/1.2179072,doi:10.1063/1.2176608,doi:10.1063/1.2387954,doi:10.1063/1.2566637,doi:10.1021/cr200107z}, which has been developed to understand the 
collective spurious self-interaction of approximated electron
densities.
In a system with a continuously  variable occupation number, 
many-body SIE may be defined  
as the deviation  from piecewise linearity of the approximate DFT total-energy with respect to the total electron  count~\cite{PhysRevLett.49.1691}.

The SIE is  most problematic for systems 
comprising spatially localized, partially filled frontier 
orbitals including those of $1s$ and $2p$ but
more canonically $3d$ and $4f$ character, 
where the qualitative failure of local and semi-local functionals has been thoroughly 
analysed~\cite{Austin71,PhysRevB.30.4734,PhysRevB.44.943,PhysRevB.71.035105}.
First-row transition metals systems thus 
can often 
benefit from corrective measures that augment conventional 
closed-form density functionals.
An approach that is very widely used  at present
is the computationally expedient DFT+$U$, 
which has been successfully applied to both extended solids~\cite{Austin71,PhysRevB.44.943,PhysRevB.48.16929,PhysRevB.50.16861,PhysRevB.52.R5467,0953-8984-9-4-002,PhysRevB.58.1201,PhysRevB.84.115108} and molecular systems~\cite{doi:10.1021/jp070549l,PhysRevB.82.081102,doi:10.1063/1.3489110,doi:10.1063/1.3660353,doi:10.1021/jz3004188,QUA:QUA24521} alike. 

DFT+$U$ attains the status of a first-principles 
method through the direct
calculation of the requisite Hubbard $U$ parameters, 
and for which a number of methods 
have been proposed~\cite{PhysRevB.58.1201,PhysRevB.71.035105,PhysRevLett.97.103001,PhysRevB.74.125106,PhysRevB.83.121101,QUA:QUA24521}. 
We refer the reader to Ref.~\citenum{linscott2018role} for a recent detailed
analysis of Hubbard $U$ and Hund's $J$ (the analogous quantity for
quantifying erroneous energy-magnetization curvature in approximate DFT)
calculation in the case of open-shell systems.
DFT+$U$ is compatible  with linear-scaling 
methods~\cite{PhysRevB.73.045110,PhysRevB.85.085107} 
intended for spatially complex systems, as well as with high-throughput materials
discovery approaches~\cite{nmat3568,PhysRevX.5.011006}.
Beginning with Ref.~\onlinecite{PhysRevLett.97.103001}, 
and continued in
Refs.~\onlinecite{doi:10.1063/1.3489110,PhysRevB.84.115108,doi:10.1063/1.3660353,:/content/aip/journal/jcp/145/5/10.1063/1.4959882}, 
the concept of DFT+$U$ as a corrective method
for SIE has been extensively developed, 
with the Hubbard $U$ parameters playing the role of localized 
error quantifiers of SIE for the approximate 
functional applied to the specific system at hand~\cite{PhysRevB.71.035105}.
We invoke this interpretation in what follows. %

\subsection{Self-interaction error in the excited-state regime}

For the integer-occupancy systems  routinely simulated, 
the generalized Koopman's condition~\cite{PhysRevB.82.115121}
gives a unified, practicable expression for the SIE-free 
condition, the non-compliance with which is, in most cases, 
responsible for the underestimated insulating 
gaps~\cite{PhysRevB.77.115123,doi:10.1021/cr200107z}
emblematic of practical DFT.
When this cannot obviously be enforced,  however, such as in neutral excited states, 
it will be helpful to decompose SIE into two contributions.
The first is an overestimation of the net self-repulsion
of the electron density due to the spurious self-interaction of
individual electron densities, particularly so for localized 
atomic orbitals, which gives rise to a positive
energy-occupancy curvature, over-delocalised of densities, 
and inaccurate ground-state total energies.
The second is the lack of any distinction 
between the density due to electrons already existing 
in a system and that due to any newly removed or added electrons,
which results in the spurious absence of derivative
discontinuities in the energy-occupancy curve
and, consequently, the shallowing of electron removal and addition levels
and the underestimation of insulating gaps.
Adiabatic linear-response  TDDFT inherits 
both components of SIE from the underlying approximate
DFT functional. 
In this work, we will focus on the former
component while treating the latter 
only at the level available
within first-principles DFT+$U$.
Technically, we use DFT+$U$ in its 
 simplified rotationally-invariant formalism 
 (which does 
 not introduce a derivative discontinuity
 but emulates the effects of one in the
Kohn-Sham~\cite{PhysRev.140.A1133} eigenspectrum), 
  with first-principles  
  linear-response Hubbard $U$ and 
 Hund's $J$ parameters.

The effect of SIE on electron dynamics 
and neutral electronic excitations,  such as those
routinely studied using TDDFT, has
slowly attracted increasing investigation
in recent years~\cite{0953-4075-31-9-006,PhysRevLett.101.096404,PhysRevLett.108.146401,doi:10.1063/1.4742763}. 
It is  a matter of central importance, for
example, in the first-principles simulation of out-of-equilibrium nanoscale 
functionalities such as dynamical
Coulomb blockade~\cite{PhysRevB.86.201109,PhysRevB.88.241102}, 
and in the first-principles  spectroscopy of systems comprising 
transition-metal ions~\cite{doi:10.1021/ic020580w,
doi:10.1063/1.3668085,doi:10.1021/jp3095227,
doi:10.1021/ct500787x,doi:10.1021/ct4010273,C5RA12962A}.
In the realm of non-atomistic calculations, the TDDFT 
solution of Hubbard type models have also attracted 
attention~\cite{PhysRevLett.101.166401,PhysRevB.90.195149,
C4CP00118D,PhysRevB.79.195127}, 
and TDDFT has also been combined 
with dynamical 
mean-field theory~\cite{PhysRevLett.106.116401,computation4030034}.

\subsection{Motivation: 
Hubbard $U$ correction in the excited-state regime of TDDFT}

Somewhat surprisingly, perhaps, given its relatively
moderate computational cost  and conceptual simplicity, 
the error correction of approximate
TDDFT by means of DFT+$U$, in the guise of
adiabatic TDDFT+$U$, has received relatively little attention
to date.
TDDFT+$U$ is readily compatible with linear-scaling  DFT, as demonstrated in the present work though the combination of linear-scaling DFT+$U$~\cite{PhysRevB.85.085107,
PhysRevB.73.045110} and linear-scaling TDDFT~\cite{doi:10.1063/1.4817330,doi:10.1063/1.4936280,doi:10.1021/acs.jctc.5b01014}, as well as with high-throughput materials  screening techniques, where DFT+$U$ is commonplace~\cite{nmat3568}. 
Within its range of applicability, 
TDDFT+$U$ could potentially offer substantial 
efficiency advantages over more involved 
methods for calculating neutral excitations in complex transition-metal
molecules and solids. 
These include  hybrid 
TDDFT~\cite{doi:10.1021/jp013949w,Rosa2004}
and Green's function based methods such as GW + 
Bethe-Salpeter~\cite{doi:10.1021/ct5003658}.
Recently, the optimally-tuned, range-separated hybrid functionals~\cite{doi:10.1142/9789812830586_0004,LEININGER1997151} within TDDFT have met with promising success in the prediction of optical excitations, particularly in the lowest excitations in organic molecules and third-row transition-metal coordination complexes~\cite{doi:10.1021/ct2009363,doi:10.1021/ct5000617,doi:10.1021/acs.jctc.5b00068}.
This latter approach has been not applied to any first-row transition-metal molecules yet, to our knowledge.
The role of DFT+$U$ in calculated excitation energies, 
particularly the explicit contribution from the Hubbard term,
has been explored in Ref.~\onlinecite{doi:10.1063/1.4757286}.
The first reported TDDFT+$U$ implementation was that of Ref.~\onlinecite{2009APS..MAR.S1097Q}, combining real-time propagation and a plane-wave basis, followed by Ref.~\onlinecite{PhysRevB.82.081106}, which detailed the results of a linear-response implementation applied to bulk NiO.
In that system, TDDFT+$U$ was shown to be capable of reproducing the experimentally observed, tightly-bound Frenkel excitons,
but not their multiplet structure.
These are relatively exotic spectroscopic features that neither the adiabatic LDA, nor the
random phase approximation built from LDA+$U$,
succeeded in recovering to any extent.
Recently, in Ref.~\onlinecite{doi:10.1021/acs.jctc.5b00895}, a real-time 
plane-wave TDDFT+$U$ implementation has been coupled
with Ehrenfest molecular dynamics to simulate both long and
short-ranged dynamical charge-transfer between alkali 
atom impurities and conjugated carbon systems.
This work revealed 
the tendency for an increasing Hubbard $U$ 
to promote the availability of multiple low-energy 
states in such systems, as well as to increase 
in energy and broaden the impurity-bath 
charge-transfer resonances.

To date, however, information has been lacking on how
the Hubbard $U$ correction affects the typical 
products of practical TDDFT calculations in 
simple transition-metal systems, namely
the low-energy excitation spectra and dipole-dipole absorption
spectra, for better or worse with respect to experiment.
Indeed, the precise effects of TDDFT+$U$
have yet to be systematically studied, and
its resulting range of applicability   has yet to be 
mapped out in any sense.
It is
this knowledge gap 
that we seek to
begin to fill 
with the present exploratory study.

\subsection{Outline of the paper: systematic
decomposition of the effects of Hubbard $U$ correction in  Kohn-Sham
DFT and linear-response TDDFT}

We seek to
systematically investigate the role of DFT+$U$
as it \emph{separately} alters the Kohn-Sham eigenspectrum 
underlying a linear-response TDDFT calculation, and the
TDDFT interaction kernel itself. 
For this, 
following its detailed introduction via an 
illustrative four-level
toy model in Section~\ref{section2}, 
we uncover the effects of full
TDDFT+$U$, in Section~\ref{sec:c8s1},
on two representative diamagnetic 
nickel  complexes 
(one perfectly closed-shell, one less so),
which were chosen for study  due to 
their relatively simple coordination chemistry. 
Since their Ni $3d$ sub-shells are close to being 
fully filled, nominally,
the dominant errors in the description of these molecules 
using an approximate semi-local xc-functional 
(in this work always Purdew-Burke-Ernzerhof, PBE~\cite{PhysRevLett.77.3865})
and xc-kernel (adiabatic PBE) may be 
ascribed primarily to SIE 
(electron delocalization) rather than static 
(multi-reference) correlation 
error~\cite{Cohen792,PhysRevB.77.115123}.
For these systems, in Section~\ref{section4},
 we show that 
first-principles Hubbard $U$
correction at the TDDFT level alone, leaving the 
underlying Kohn-Sham eigenspectrum at its DFT level, 
offers a far better agreement with available experimental
and quantum-chemical data, when compared to either uncorrected DFT \& TDDFT 
or consistent DFT+$U$ \& TDDFT+$U$.
Performing Hubbard $U$ correction at the DFT level alone
meanwhile, 
leaving the TDDFT kernel uncorrected, 
leads to very unreasonable results indeed.
We will discuss some implications and possible solutions
to this intriguing asymmetry in Section~\ref{Section6}.

We will turn first, however, in Section~\ref{Section5}, to the  technically challenging case of an 
open-shell system, a non-centrosymmetric, 
high-spin cobalt coordination complex.
Here, we will again find that a first-principles 
DFT+$U$ correction applied only 
to the Kohn-Sham eigenspectrum 
drastically degrades the agreement between 
the singlet excitation and the dipole-dipole absorption spectra
and, respectively,  high-level
quantum-chemical and experimental data.
The agreement is recovered to some degree when
TDDFT+$U$ is also used, but a number of important
spectral features remain poorly described.
In this case, we will show that the application of 
first-principles TDDFT+$U$ upon the
DFT Kohn-Sham eigenspectrum is not a viable
work-around, as the implied 
inconsistency leads to
the emergence of a singlet instability.

\section{Hubbard correction of the exchange-correlation kernel: theory and numerical illustration}
\label{section2}

Let us now introduce the anatomy of the
Hubbard $U$ correction to approximate TDDFT.
Concerning ourselves only with low-energy
single-particle excitations,  
we will restrict ourselves to the linear-response regime.
Here,  the spin-unpolarized TDDFT  problem may 
be  expressed in the occupied-unoccupied Kohn-Sham
eigenvector  product space via Casida's equation~\cite{Casida1996391,Casida20093}, which is an
eigen-equation for the vertical 
excitation frequencies $\omega$,  given 
in its canonical notation by 
\begin{align}\label{eq:e1}
\left( \begin{array}{cc}
\mathbf{A} & \mathbf{B} \\
\mathbf{B}^\dagger & \mathbf{A}^\dagger
\end{array}\right)
\left(\begin{array}{c}
\mathbf{X} \\
\mathbf{Y}
\end{array}\right)   
=\omega
\left(\begin{array}{c}
\mathbf{X} \\
\mathbf{-Y}
\end{array}\right).
\end{align} 
The
Hamiltonian
matrix elements 
$A_{cv, c' v'}=\delta_{v v'}\delta_{c c'}\omega_{c'v'}+K_{cv,c' v'}$ and $B_{cv, c' v'}=K_{cv, v' c' }$ correspond to 
excitation-excitation  pairs and 
excitation-relaxation pairs, respectively. 
The neglect of coupling between these
processes, that is the approximation $\mathbf{B} =
\mathbf{0}$, is known as the Tamm-Dancoff
approximation (TDA).
The ground-state Kohn-Sham 
eigenvalues $\epsilon_v$ are those of occupied valence states,
while the $\epsilon_c$ are those of unoccupied conduction states.
The coupling matrix $\textbf{K}$ incorporates all interactions
between particle-hole pairs, which is to say all effects
 beyond the many-body random-phase approximation
(Fermi's Golden Rule, or FGR).
It is given, within the valence-conduction $\left( c v \right)$ product   
 representation of the  interaction kernel $\hat{f}$, by
\begin{align}\label{eq:e2}
K_{cv,c'v'}={}&\iiiint d\mathbf{r}\;d\mathbf{r}'\;d\mathbf{r}''\;d\mathbf{r}'''\;\psi_{c}^*\left(\mathbf{r}\right)\psi_{v}\left(\mathbf{r}'\right) \\ 
& \quad \times f\left(\mathbf{r},\mathbf{r}',\mathbf{r}'',\mathbf{r}'''\right)
\psi_{c'}\left(\mathbf{r}''\right)
\psi_{v'}^*\left(\mathbf{r}'''\right)\nonumber,
\end{align}
where the 
$\psi$ are Kohn-Sham eigenvectors.
The kernel ordinarily comprises  Hartree  and xc 
terms only, denoted by $\hat{f}_\textrm{H}$
and $\hat{f}_\textrm{xc}$, 
but if  a DFT+$U$ derived correction term $\hat{f}_U$ is added, 
the resulting TDDFT+$U$ interaction kernel is given by
$\hat{f}=\hat{f}_U + 2 (
\hat{f}_\textrm{H}+\hat{f}_\textrm{xc})$. 
The underlying Kohn-Sham eigensystem is also  changed,  typically.
The factor of $2$ here is conventional,
and it represents the sum of  
identical (in the unpolarized case)
like and unlike-spin Hartree and xc interactions
acting on a given excitation.
This factor of $2$ does not, however,  pre-multiply 
$\hat{f}_U$, since  DFT+$U$ ordinarily acts explicitly 
only on like-spin Kohn-Sham states.
The rotationally-invariant DFT+$U$ energy
functional~\cite{PhysRevB.44.943,PhysRevB.48.16929,PhysRevB.50.16861,PhysRevB.52.R5467,0953-8984-9-4-002}
used in this work falls into this category, 
being given, for a SIE-affected subspace, by
\begin{align}\label{eq:e3} 
&  E_U  = \frac{U_\mathrm{eff}}{2}\sum_{\sigma}\sum_m \left(n^{\sigma}_{mm}-\sum_{m'}n^{\sigma}_{mm'}n^{\sigma}_{m'm}\right),
\end{align}
where $U_\mathrm{eff}=U -J$ is the effective like-spin correction parameter expressed in terms of the Hubbard $U$ and the  Hund's  $J$ parameter. The index $\sigma$ is for spin, and the subspace occupancy matrix 
 $n^\sigma_{m m'} = \sum_v \langle 
 \varphi_m
 \rvert \psi^\sigma_v \rangle  
 \langle \psi^\sigma_v \lvert \varphi_{m'}\rangle $
is typically defined in terms of localized 
orbitals (in our calculations, orthonormal 
atomic nickel or cobalt $3d$ orbitals solved
in a norm-conserving pseudopotential), $\varphi_m$.
The Hubbard $U$ \emph{kernel} 
is the second functional derivative~\cite{PhysRevLett.52.997}
 of the DFT+$U$ energy $E_U$ with respect to the 
 density matrix, and  we find,
denoting the density-matrix for  spin $\sigma$ by 
by $\rho^\sigma \left(\mathbf{r},\mathbf{r}'\right)$, that
\begin{align}\label{eq:e4}
f^{\sigma \sigma'}_U & \left(\mathbf{r},\mathbf{r}',\mathbf{r}'',\mathbf{r}'''\right) =
\frac{\delta^2 E_U[\rho^\sigma,\rho^{\sigma '}]}{\delta \rho^\sigma \left(\mathbf{r}'',\mathbf{r}'''\right)\delta 
\rho^{\sigma '} \left(\mathbf{r},\mathbf{r}'\right)} \\
 \nonumber
&= -U_\mathrm{eff} \sum_{mm'}
\delta^{\sigma \sigma'}
\varphi_m\left(\textbf{r}\right)\varphi_{m'}^*\left(\textbf{r}'\right)
\varphi^*_m\left(\textbf{r}''\right)\varphi_{m'}\left(\textbf{r}'''\right).
\end{align}
The resulting Hubbard $U$ contribution  to 
$\textbf{K}$ may be  written, using implicit 
summation of paired indices, as
\begin{align}\label{eq:e5}
K_{cv,c'v'}^U ={}& -U_\textrm{eff}
\langle \psi_c|\varphi_m\rangle 
\langle\varphi_{m'}|\psi_v\rangle 
\left( 
\langle \psi_{c'} | \varphi_m \rangle 
\langle \varphi_{m'}| \psi_{v'}\rangle  
\right)^\ast
\nonumber \\
={}& -U_\textrm{eff}
\langle\psi_c |\varphi_m\rangle 
\langle\varphi_{m}|\psi_{c'}\rangle
\nonumber \\ 
{}&  \times
 \langle \psi_{v'}|\varphi_{m'}\rangle 
\langle\varphi_{m'}|\psi_{v}\rangle, 
\end{align}
whereafter we will use $U$ rather $U_\textrm{eff}$ for simplicity, except where discussing our actual calculated
$U_\textrm{eff}$.
The resulting `direct' term, in what can be seen as an effective 
exciton self-interaction 
correction, is given by
\begin{equation}
K_{cv,cv}^U = - U \sum_{m m'}
\lvert  \langle \psi_c \rvert \varphi_m \rangle \rvert^2
\lvert  \langle \psi_v \rvert \varphi_{m'} \rangle \rvert^2.
\end{equation}

The form of $\mathbf{K}^U$  hints at the  behaviour expected 
of  the TDDFT+$U$ excitation spectrum as  $U$ is varied.
For $U > 0$~eV, the interaction correction 
due to one $\left(   c  v \right)$ pair and acting upon another 
is a sum over (typically) attractive direct Hartree and 
exchange terms. 
Relative to the situation that holds in hybrid-exchange TDDFT, however, 
the exchange terms are expected to be more significant relative to 
 direct Hartree ones, 
 since in TDDFT+$U$ the same constant $U$  pre-multiplies  both 
 term types. 
It is instructive to examine the special case in which
the projecting orbitals $\varphi_m$ are identical to a subset of
the underlying Kohn-Sham states $\psi$. 
There, the Hubbard $U$ contributions to \textbf{B} and \textbf{A} 
reduce considerably to 
\begin{align}\label{eq:e6}
A_{cv,c'v'}^U=&-U\delta_{cm}\delta_{mc'}
\delta_{v'm'}\delta_{m'v}=-U\delta_{cc'}\delta_{vv'}, \hspace{0.5cm}\mathrm{and} \nonumber \\ 
B_{cv,c'v'}^U=&-U\delta_{cm}\delta_{mv'}
\delta_{c'm'}\delta_{m'v}=0,
\end{align}
leaving a fully diagonal contribution to the Casida Hamiltonian.
If these Kohn-Sham states are also well separated from all others
energetically, the effect of the Hubbard $U$ on the underlying
eigenstate differences $\epsilon_c - \epsilon_v $ will simply be an
increase by $U$, whereupon the effects of DFT+$U$
 and TDDFT+$U$ \emph{fully cancel} for excitations coupling states
 within the target subspace.
This picture is  complicated 
 by Kohn-Sham state hybridization, self-consistency, and the spillage of 
 the localized orbitals, in practice.  
 Nonetheless, the TDDFT+$U$ correction may  be expected 
 to increase the mixing of transitions between states 
 that overlap strongly with the selected subspace, 
 and to increase their exciton binding energy by 
compensating for the underlying DFT+$U$ eigenvalue correction.
  However, the matrix elements of $\mathbf{K}^U$ are
 quadratic in overlap integrals of the form
 $\langle\psi_c |\varphi\rangle 
 \langle\varphi|\psi_c\rangle$,
 whereas  the underlying Hubbard $U$ correction to the Kohn-Sham
 potential comprises terms that are only linear in such integrals.
Thus, we cannot generally expect the cancellation of
the $U$ correction to the ground and excited-state
systems to be precise in practical calculations.

\subsection{Illustration of the effect of $U$ correction in TDDFT using a four-level toy model}

\begin{figure}[h]
\resizebox{0.45\textwidth}{!}{
\begin{tikzpicture}
    \node [rectangle,text width=1em, rounded corners, minimum height=2em,line width=0.5mm] (c2) {c'};
    
    \node [rectangle, text width=5em, rounded corners, minimum height=2em,line width=0.5mm, right of=c2,node distance=6cm] (ec2) {$\epsilon_{c'}$ = 6 eV};
    
    \node [rectangle, text width=1em,  rounded corners, minimum height=2em,line width=0.5mm, below of=c2,node distance=0.75cm] (c1) {c};
    
    \node [rectangle, text width=5em,  rounded corners, minimum height=2em,line width=0.5mm, right of=c1,node distance=6cm] (ec1) {$\epsilon_{c}$ = 2 eV};

    \node [rectangle, text width=1em,  rounded corners, minimum height=2em,line width=0.5mm, below of=c1,node distance=1cm] (v1) {v};
    
    \node [rectangle, text width=5em, rounded corners, minimum height=2em,line width=0.5mm, right of=v1,node distance=6cm] (ev1) {$\epsilon_{v}$ = -2 eV};
    
    \node [rectangle, text width=1em, rounded corners, minimum height=2em,line width=0.5mm, below of=v1,node distance=1.5cm] (v2) {v'};
    
    \node [rectangle, text width=5em,  rounded corners, minimum height=2em,line width=0.5mm, right of=v2,node distance=6cm] (ev2) {$\epsilon_{v'}$ = -8 eV};        
        
    
    \path [draw,line width=0.1cm] (c2) -- (ec2);
    \path [draw,red,dashed,line width=0.1cm] (c1) -- (ec1);
    \path [draw,red,dashed,line width=0.1cm] (v1) -- (ev1);
    \path [draw,line width=0.1cm] (v2) -- (ev2);
    \draw [line width=0.075cm] (-1,-4) -- (-1,1) --(8,1) -- (8,-4) -- (-1,-4);
\end{tikzpicture}}
\caption{
A four-level toy model for independent-particle (Kohn-Sham orbital emulating) states of  arbitrarily assigned energies, 
comprising two levels affected by $U$ corrections and illustrated with  dashed red lines and two bystander levels illustrated with the black lines.}
\label{fig:toymodel-ill}
\end{figure}
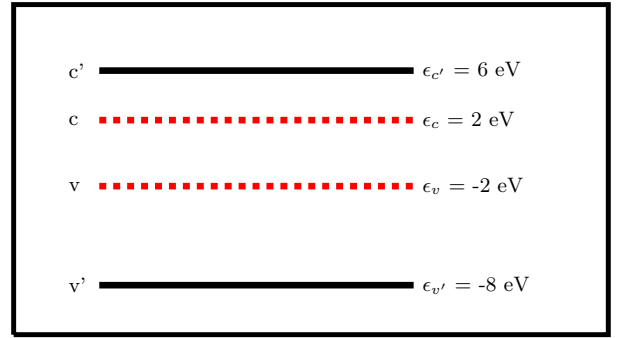

For further insight, the effects of \uu in conjunction with \u can be illustrated by means of a toy model in conjunction with  the TDA and  full Casida equation.
Let us consider  four independent-particle (KS-like) states,  of which two occupied and two unoccupied states are labelled with $\{v,v'\}$ and $\{c,c'\}$,  respectively, with some arbitrary eigenenergies as illustrated in Fig.~\ref{fig:toymodel-ill}.
The pair $\{c,v\}$ of states shown in dashed-red are targeted with a correction inspired by \u and TDDFT+$U$.

The block matrices $\mathbf{A}$ and $\mathbf{B}$ in the Casida equation become $4\times 4$  matrices with elements
given by 
\begin{align}\label{eq:c7e71}
 A_{ji,j'i'}=&\Big[(\epsilon_{j'}-\epsilon_{i'})+\frac{U_\textrm{DFT}}{2}\left(\delta_{j'c}+\delta_{i'v}\right)  \\
& \quad - U_\textrm{TDDFT}\delta_{j'i',cv} \Big] \delta_{i'i}\delta_{jj'} 
+K^{\mathrm{Hxc}}_{ji,j'i'},
\nonumber \\
B_{ji,j'i'}=&K^{\mathrm{Hxc}}_{ji,i'j'}
\end{align} 
where $j$ and $j'$ run over $\{c,c'\}$, while $i$ and $i'$ run over $\{v,v'\}$. 
The  Hubbard parameter $U_\textrm{DFT}$ imitates the  effect of \u by pushing the targeted states away from the Fermi level via the term $U_\textrm{DFT} \left(\delta_{j'c}+\delta_{i'v} \right) /2$, whereas the   Hubbard parameter $U_\textrm{TDDFT}$ includes the effect of \uu via the term $-U_\textrm{TDDFT} \delta_{j'i',cv}$. 
By making these two Hubbard parameters  $U_\textrm{DFT}$ and $U_\textrm{TDDFT}$ independent, the individual effects of the Hubbard corrections at the  DFT and TDDFT levels can be observed by setting one of them to zero at a time.
The Hartree+xc coupling matrix elements are assigned for illustration here 
to  the arbitrary values 
\begin{align}\label{eq:c7e72}
& K^{\mathrm{Hxc}}_{ji,j'i'} =
\begin{cases}
4.0 \; eV& \mathrm{for} \hspace{0.5cm} \delta_{ji,j'i'}   \\
0.8 \; eV & \mathrm{ otherwise,}
\end{cases} \nonumber \\
&K^{\mathrm{Hxc}}_{ji,i'j'} =
\begin{cases}
4.0 \; eV& \mathrm{for} \hspace{0.5cm} \delta_{ji,i'j'}   \\
0.8 \; eV &  \mathrm{otherwise},
\end{cases} 
\end{align}
and the symmetric choice made here
is a deliberate attempt to simplify the contributions due to $f_{\mathrm{Hxc}}$.

The Casida equation, both in its full form and within the TDA, was solved using an eigenvalue solver over a range of
$U_\textrm{DFT}$ and $U_\textrm{TDDFT}$ values. 
Additionally, FGR excitations energies are included and calculated  as
\begin{align}\label{eq:c7e73}
\omega^\mathrm{FGR}_{ji}(U_\textrm{DFT})=(\epsilon_{j}-\epsilon_{i})+\frac{U_\textrm{DFT}}{2}\left(\delta_{jc}+\delta_{iv}\right).
\end{align}
\begin{figure}[t!]
\centering
 \includegraphics[width=0.2\textwidth]{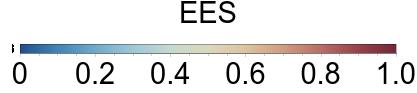}

\subfloat[\u \& FGR \label{fig:toy-tda-fgr}]{
\includegraphics[width=0.22\textwidth]{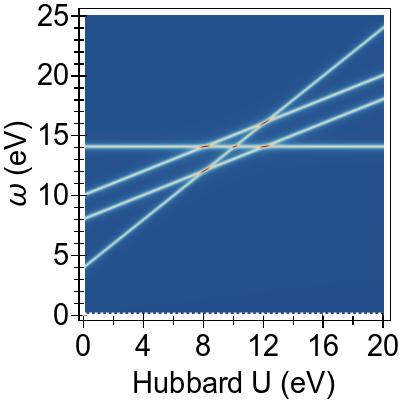}
}\hfill
\subfloat[\u \& TDDFT  \label{fig:toy-rpa-uuo}]{
\includegraphics[width=0.22\textwidth]{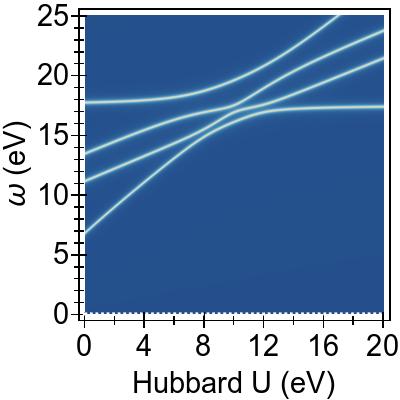}
}

\subfloat[DFT  \& \uu  \label{fig:toy-rpa-oou}]{
\includegraphics[width=0.22\textwidth]{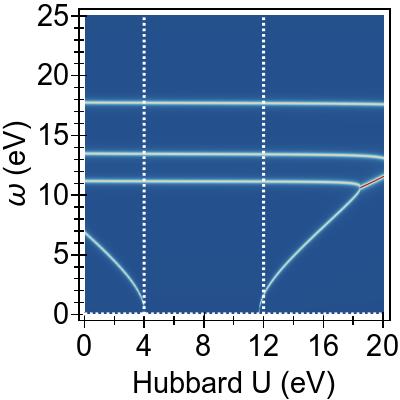}
}\hfill
\subfloat[\u \& \uu  \label{fig:toy-rpa-uuu}]{
\includegraphics[width=0.22\textwidth]{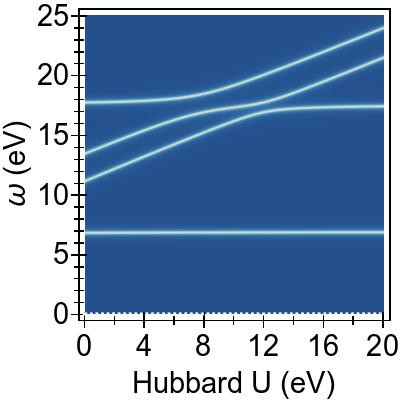}
}

\subfloat[DFT \& \uu (TDA)  \label{fig:toy-tda-oou}]{
\includegraphics[width=0.22\textwidth]{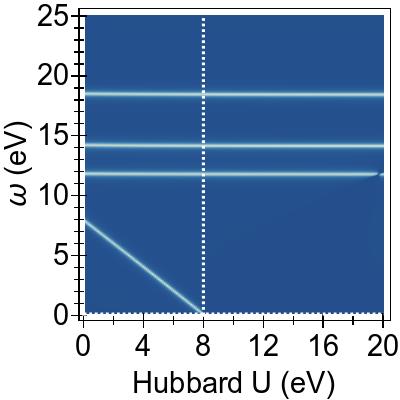}
}\hfill
\subfloat[\u \& \uu (TDA)  \label{fig:toy-tda-uuu}]{
\includegraphics[width=0.22\textwidth]{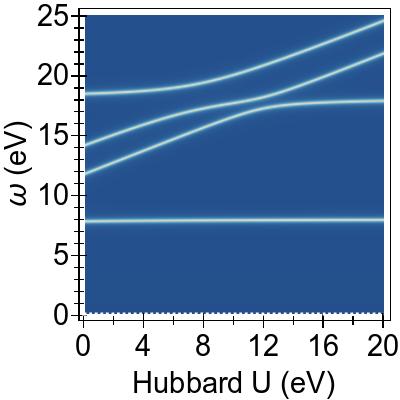}
}

\caption{Electronic excitation spectra (EES) calculated from our illustrative 
four-level toy model, 
using Eq.~\eqref{eq:c1e38} with $\Gamma=0.1$~eV. 
Sub-figure captions indicate the analogous DFT-based approximation, e.g., the +$U$ in `DFT+$U$ \& FGR' indicates that the occupied (unoccupied) localised level is lowered (raised) by $U_\textrm{DFT}/2$ (with $U_\textrm{TDDFT} = 0$~eV), while `FGR' indicates that the transitions are treated as independent. On the other hand, `TDDFT' denotes that a repulsive kernel given by Eq.~\eqref{eq:c7e72} couples transitions, while `TDDFT+$U$' indicates that  said kernel is $U$-corrected by Eq.~\eqref{eq:c7e71} with the Hubbard $U$ axis denoting $U_\textrm{TDDFT}$. For `DFT+$U$ \& TDDFT+$U$', 
$U_\textrm{DFT} = U_\textrm{TDDFT}$.
TDA is the Tamm-Dancoff approximation.}
\label{fig:toy}
\end{figure}

In Fig.~\ref{fig:toy}, the principal effects of a positive $U_\textrm{DFT}$
(simulating DFT+$U$) and $U_\textrm{TDDFT}$ (simulating TDDFT+$U$)
in our toy model 
are demonstrated,  via the amplitudes of normalised 
electronic excitation spectra
(EES) calculated using Eq.~\eqref{eq:c1e38}. 
A life-time broadening of $\Gamma = 0.1$~eV 
was used here, together with a  high-resolution  grid of  Hubbard $U$ parameters taken in $0.05$~eV  steps. 
Starting from the energy levels shown in Fig.~\ref{fig:toymodel-ill},  a positive value of $U = U_\textrm{DFT}$ pushes the targeted  (red-dashed in Fig.~\ref{fig:toymodel-ill}) states $(v,c)$ away from the Fermi level, each by with $U/2$, while the bystander states remain intact. 
Consequently,  in Fig.~\ref{fig:toy-tda-fgr},    the excitation from  $v$ to $c$ ($v\rightarrow c$) increases simply by $U$, while  the energies of $v'\rightarrow c$ and $v \rightarrow c'$ increase by $U/2$, emulating the effects of DFT+$U$. 
The remaining excitation $v'\rightarrow c'$  is not affected due to lack of interaction between exciton pairs within FGR.

Comparing next Figs.~\ref{fig:toy-rpa-uuo},~\ref{fig:toy-rpa-oou}, and~\ref{fig:toy-rpa-uuu} against the FGR results of  Fig.~\ref{fig:toy-tda-fgr}, taken each at $U=0$~eV, a global shift 
by TDDFT of $\sim 3-4$~eV  on the excitation energies can be seen, as
well as the avoided crossing of excitation energies for $U > 0$~eV.
This is due to the interactions between exciton pairs, 
emulating TDDFT, that are introduced by the coupling matrix  $K^\mathrm{Hxc}_{ji,j'i'}$ in Eq.~\eqref{eq:c7e72}.
The global nature of the shift is due to the invariance of the coupling matrix with respect to the swapping of orbital indices.
In Figs.~\ref{fig:toy-rpa-oou} and ~\ref{fig:toy-tda-oou}, 
the $U_\textrm{TDDFT}$ term (emulating TDDFT+$U$) exclusively affects the excitation $v\rightarrow c$ by  pushing it down 
(linearly in the TDA case)
from $\approx 8$~eV for increasing $U = U_\textrm{TDDFT}$ values. 
For $U=4$~eV ($U=8$~eV for TDA), the excitation $v\rightarrow c$ becomes purely imaginary
(negative in the TDA case),
meaning that the model becomes unphysical.
In Fig.~\ref{fig:toy-rpa-uuu}, the combined emulated
effects of \u and TDDFT+$U$, when $U_\textrm{DFT} = U = U_\textrm{TDDFT}$, are seen in the form of a total cancellation of  the effect of \u on the excitation $v \rightarrow c$ by TDDFT+$U$.
 The remaining three excitations are affected by DFT+$U$ 
 as before, while the effect of TDDFT+$U$ (comparing 
 Figs.~\ref{fig:toy-rpa-uuo} and ~\ref{fig:toy-rpa-uuu})
 is relatively minor and mostly due to avoided crossing.
Comparing Fig.~\ref{fig:toy-rpa-uuu} with its TDA counterpart
Fig.~\ref{fig:toy-tda-uuu}, the excitations within this model show a similar 
qualitative behaviour 
irrespective of whether the
 TDA is invoked.
 The TDA approximately shifts the excitations up in energy by 
 $\sim 1$ eV  throughout the frequency range.
 
\subsection{Implementation of the TDDFT+\textit{U} kernel
within linear-scaling linear-response TDDFT}

We have implemented the TDDFT+$U$ kernel of 
Eq.~\ref{eq:e4} in the \textsc{ONETEP} 
package~\cite{doi:10.1063/1.1839852,PSSB:PSSB200541457,
PhysRevB.85.085107}.
This direct-minimization DFT
code maintains a linear-scaling increase in computational
expense with respect to system size, while maintaining
an accuracy which is effectively 
equivalent to that of a plane-wave code.
It does this  by expanding the Kohn-Sham density-matrix in terms
of a minimal set of spatially truncated
non-orthogonal generalized Wannier functions (NGWFs), 
which are variationally optimized \emph{in situ}~\cite{PhysRevB.66.035119}.
For calculations involving excited states, 
the code is capable of variationally optimizing a 
set of Wannier functions for the unoccupied conduction bands
as a post-processing step that follows conventional total-energy 
minimization~\cite{PhysRevB.84.165131}.
With this, and using the resulting joint basis of optimized
valence and conduction band Wannier functions,
we used the linear-scaling beyond-Tamm-Dancoff linear-response 
TDDFT functionality
available in  \textsc{ONETEP}
~\cite{doi:10.1063/1.4817330,doi:10.1063/1.4936280,doi:10.1021/acs.jctc.5b01014}, which again uses iterative minimization,
as the basis for our implementation.
The central element in our combination of linear-scaling TDDFT
and DFT+$U$~\cite{PhysRevB.85.085107} is the 
\emph{change} in DFT+$U$ potential associated with the 
first-order change in Kohn-Sham density-matrix,
$\rho^{ \left( 1 \right) } ( \mathbf{r}, 
\mathbf{r'}; \omega)$ at a each excitation energy $\omega$, 
which is given by the same expression
for both singlet and triplet excitations alike, specifically
\begin{align}
 \hat{V}^{\sigma \left( 1 \right) }_U \left( \omega \right) 
 = - U \sum_{m m'} \lvert \varphi_m \rangle \langle \varphi_m \rvert
 \rho^{\sigma \left( 1 \right) } (\omega )
 \lvert \varphi_{m'} \rangle \langle \varphi_{m'} \rvert .
\end{align}

From this equation, it is clear that the occupancy dependence of the
DFT+$U$ potential survives in TDDFT+$U$, insofar
as that, for $U > 0$~eV, 
a level within the target subspace that is depopulated
under excitation (typically a valence level close to the gap) 
will be subject to  a more repulsive DFT+$U$
potential, whereas a repopulated (e.g., conduction)
level will be subject to a more attractive
DFT+$U$ potential. 
TDDFT+$U$ thus tends to promote such excitations by
increasing  the exciton binding between the associated levels.
We emphasise that the interaction
in TDDFT+$U$ remains entirely adiabatic 
as it is presented here, since the kernel
$\hat{f}_U$ is constant, and so it addresses 
only the time-average of the self-interaction error as it is 
measured in the ground-state.
As a result, it lacks the ability to produce dynamical step
features in the potential that may result of occupancies
passing through integer values, which are  dynamical
manifestations of the second aspect of self-interaction error
previously discussed.
However, TDDFT+$U$ does provide a convenient framework
in which to explore non-adiabatic self-interaction 
correction kernels $\hat{f}_U ( \omega ) $, 
either by means of an explicitly frequency-dependent Hubbard 
$U ( \omega ) $.

\section{The Hubbard \textit{U} dependence of 
neutral excitation spectra}\label{sec:c8s1}
\begin{figure}[t]
\centering
\subfloat[\label{fig:nn-str}]{\tcbox[sharp corners, boxsep=0.0mm, boxrule=0.5mm,  colframe=gray, colback=white]{\includegraphics[width=0.15\textwidth]{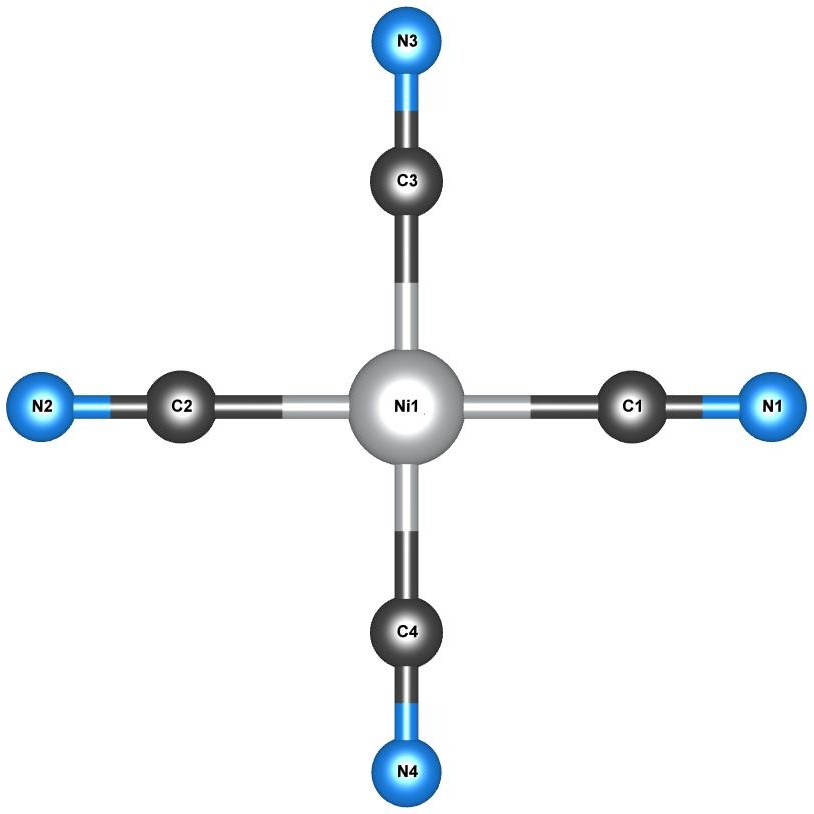}}}\hfill
\subfloat[\label{fig:nc-str}]{\tcbox[sharp corners, boxsep=0.0mm, boxrule=0.5mm,  colframe=gray, colback=white]{\includegraphics[width=0.15\textwidth]{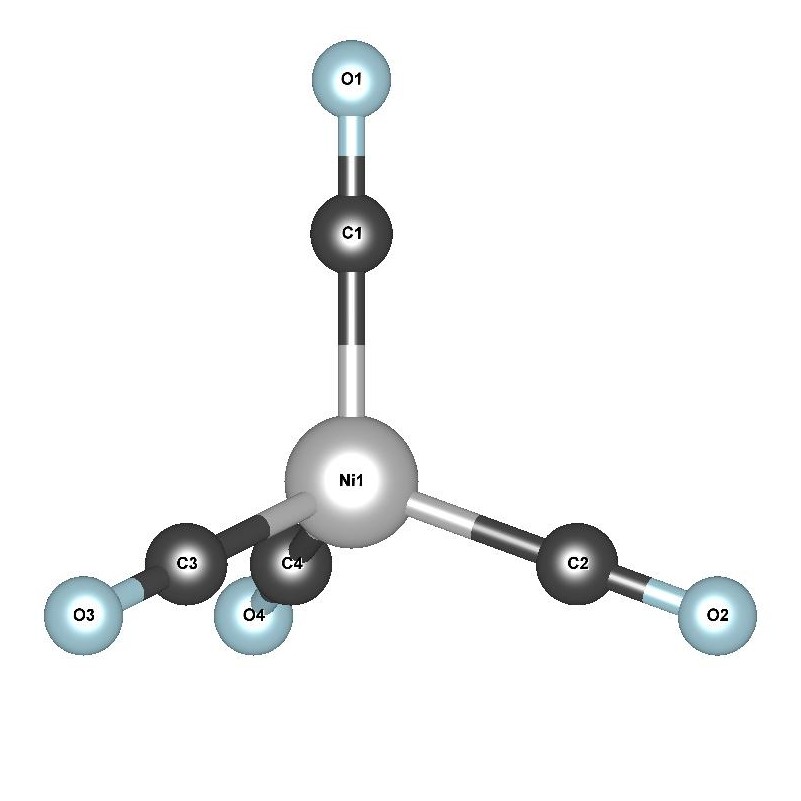}}}

\caption{The molecular structures of two representative  Ni-centered closed-shell coordination complexes. Shown
left is the 
planar tetracyanonickelate anion Ni(CN)${_4^{2-}}$, 
and shown right is the tetrahedral nickel tetracarbonyl Ni(CO)${_4}$.}
\label{fig:ni-str}
\end{figure}

Two small closed-shell Ni-centred coordination complexes,  namely the planar tetracyanonickelate anion Ni(CN)\td{4}\tu{2-}
and  tetrahedral nickel tetracarbonyl Ni(CO)\td{4} shown in Fig.~\ref{fig:ni-str}, were chosen for study. 
The Hubbard $U$ dependence of molecular spectra, in 
terms of both its individual  effects on \u and TDDFT+$U$, 
and on their combination, 
was investigated.
These systems provide a useful playground in which to investigate the effects of \u and TDDFT+$U$, since they minimise any complex contributions from  magnetic ordering and large ligand-field splittings,  as both systems are closed-shell and centro-symmetric with  strong ligands.
Furthermore, these systems have  previously been studied experimentally~\cite{doi:10.1021/ja00886a002,doi:10.1021/ja01015a013,doi:10.1021/ic50198a007,doi:10.1021/ja00202a004} and using numerous first-principles methods~\cite{doi:10.1021/jp9614355,doi:10.1021/jp991060y,doi:10.1021/ic060584r,doi:10.1063/1.4905124}. 
This is not, however, to imply that these systems are
ideal candidates for treatment using DFT+$U$, let
alone TDDFT+$U$, since they are reasonably well
described by conventional approximate DFT.

\subsubsection*{Convention for visualising spectra}
At this juncture we must introduce 
our conventions for visualising two 
essential molecular spectroscopies.
Electronic excitation spectra (EES) are constructed here by including both optically allowed and forbidden excitations with the same unit oscillator strength. They are calculated using the formula
\begin{align}\label{eq:c1e38}
\mathrm{EES}(\omega)=\sum_{ij}\frac{\Gamma/2}{\left(\omega-\omega_{ji}\right)^2+\left(\Gamma/2\right)^2},
\end{align}
where $\omega_{ji}$ denotes the energy of a transition
from an occupied ($i$) to an  unoccupied 
($j$) molecular electronic state, and
$\Gamma$ is a Lorentzian broadening factor.

Electric dipole-dipole absorption spectra are 
commonly 
used to measure the optical response of molecules in the low-energy spectral range.
The contributions of the individual excitations are weighted by  oscillator strengths
$f_{j\leftarrow i}$
 related to the transition dipole moments.
The  formula relevant to optical absorption is
\begin{align}\label{eq:c1e39}
\mathrm{ABS}(\omega)=\sum_{ij}f_{j\leftarrow i}\frac{\Gamma/2}{\left(\omega-\omega_{ji}\right)^2+\left(\Gamma/2\right)^2},
\end{align}
and this type of spectrum is 
the one primarily used here for comparing with experimental observations.

EES and ABS were constructed  using Eq.~\eqref{eq:c1e38} and Eq.~\eqref{eq:c1e39} with a Lorentzian broadening $\Gamma=0.1$~eV at  integer values of the Hubbard $U$ parameters, and interpolated to   intermediate values in $0.01$~eV steps.
%
%
Our EES are scaled by setting  the global maximum of  EES data across DFT \& TDDFT, \u\& TDDFT, DFT \& TDDFT+$U$, and \u\& TDDFT+$U$
to unity.
Similarly, our ABS are scaled by setting the global maximum of  ABS data across all of these four combinations to unity.
Such separate scaling factors enable us to compare relative intensities  within various methods as well as to maintain the comparability between EES and ABS within same method. 
EES calculated within the FGR are scaled separately, using their own maxima.

\subsection{The square-planar tetracyanonickelate anion: Ni(CN)$\mathbf{_4^{2-}}$}\label{sec:c8s2}

\begin{figure}[h]
\centering
\hfill \includegraphics[width=0.2\textwidth]{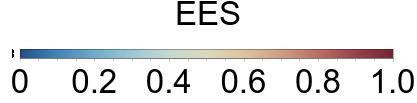}

\subfloat[Energy levels \label{fig:nn-band}]{\includegraphics[width=0.23\textwidth]{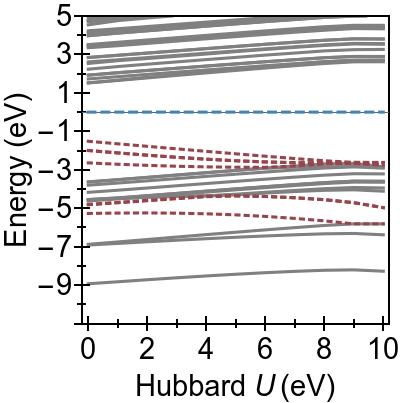}}\hfill
\subfloat[FGR \label{fig:nn-fgr}]{\includegraphics[width=0.22\textwidth]{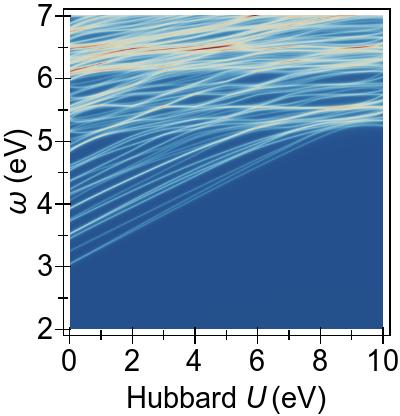}}
\caption{The Kohn-Sham DFT+$U$ energy levels and  singlet EES spectra 
of Ni(CN)$_4^{2-}$
calculated using the FGR, as  functions of  Hubbard $U$ parameter. The Fermi level (dashed, blue line)  is set to 0 eV. The states most strongly 
affected by \u are shown with dashed, red lines.}
\label{fig:nn-initial}
\end{figure}

The square-planar Ni(CN)$_4^{2-}$ is a low-spin coordination complex, with a  Ni center of  nominal charge  $2+$. 
(CN)\tu{-} is a strong-field $\pi$-acceptor  ligand that leads to ligand-splitting at $3d$-levels of Ni,  following $d_{yz}\approx d_{xz}<d_{xy}<d_{z^2}<d_{x^2-y^2}$, where $3d^8$ electrons occupy the first four levels  and  the remaining $3d_{x^2-y^2}$ forms an   $dsp^2$-hybrid with the ligands  in the square-planar symmetry~\cite{griffith1964theory}.
As a result, the low-lying excitations are expected to be predominantly of a mixed  $3d\rightarrow 3d$  and  metal-to-ligand $3d\rightarrow \pi^*$ character,  as suggested by  previous studies~\cite{doi:10.1021/ic060584r}.

The energy alignment of $3d$ states is shown as a function of $U$ in Fig.~\ref{fig:nn-band}.
For  increasing $U$ values, the occupied $3d$ states move to deeper energies.
The states close to the HOMO-LUMO gap (shown with red, dashed lines), which  strongly contribute to  low-lying excitations, fall  to  lower energetic states entirely at about  $U\gtrapprox7$~eV.
Thus,   low-lying excitations are pushed upwards and, ultimately, they combine with higher energy excitations of metal-to-ligand character, as seen in the EES 
calculated using FGR in Fig.~\ref{fig:nn-fgr}.

\begin{figure}[t]
\centering
\includegraphics[width=0.2\textwidth]{mol-ees-legend} 

\subfloat[ \u \& TDDFT \label{fig:nn-uuo-exc}]{\includegraphics[width=0.22\textwidth]{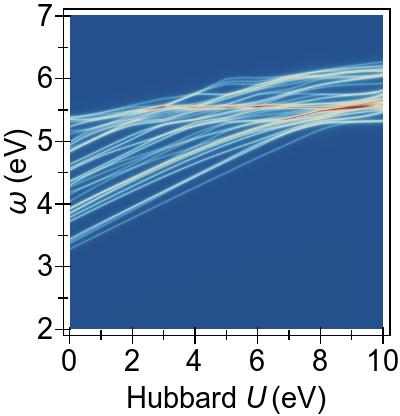}}\hfill
\subfloat[DFT  \& \uu \label{fig:nn-oou-exc}]{\includegraphics[width=0.22\textwidth]{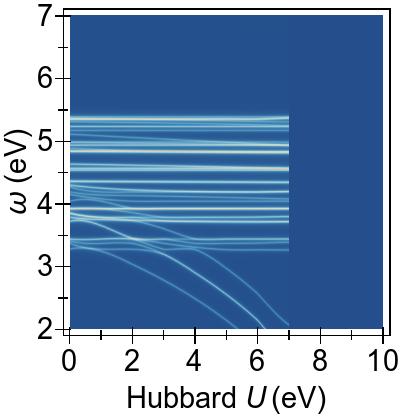}}

\subfloat[ \u \& \uu \label{fig:nn-uuu-exc}]{\includegraphics[width=0.22\textwidth]{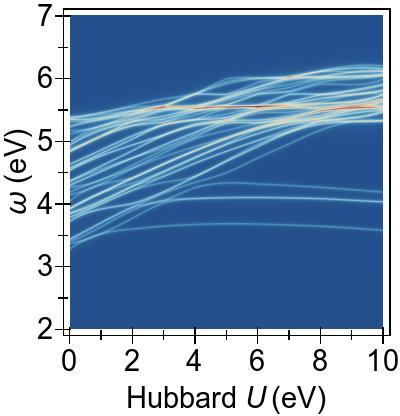}}\hfill
\subfloat[ \u \& \uu (TDA)  \label{fig:nn-uuu-exc-tda}]{\includegraphics[width=0.22\textwidth]{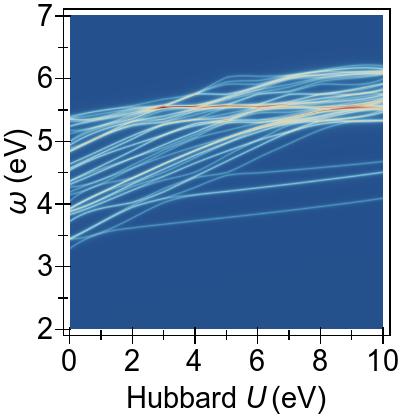}}

\caption{The singlet EES of Ni(CN)$_4^{2-}$, calculated using \u\& TDDFT, DFT \& TDDFT$+U$, and \u\& TDDFT+$U$,    as  functions of  the Hubbard $U$ parameter.}
\label{fig:nn-exc}
\end{figure}

Up to this point, the Hubbard $U$ has been used  only to modify the under-lying KS-DFT states via DFT+$U$.
In Fig.~\ref{fig:nn-exc}, a more complete and consistent picture is provided, by the EES  for the first 50 singlet excitations calculated using various combinations of \u and TDDFT+$U$. 
In Fig.~\ref{fig:nn-uuo-exc}, we see that an increasing $U$ value in \u reduces  the $3d$ $\rightarrow$ $3d$ character of the excitations, and combines them with  excitations from deeper states, similarly to the FGR case. 
Beyond that, \u is effective globally insofar as that it pushes other excitations to higher energies as well,   by means of modifying  the metal-to-ligand energy as seen in Fig.~\ref{fig:nn-band}.

On the contrary, in Fig.~\ref{fig:nn-oou-exc}, 
we observe that  \uu affects  only the excitations  of  $3d \rightarrow 3d$ character, 
while, as anticipated, the  remaining excitations remain largely unaffected. 
Furthermore, the affected excitations become non-physical for $U\gtrapprox7$~eV in DFT \& TDDFT+$U$,  similarly to what is observed in the  four-level toy model.
This situation arises by virtue of  exciton over-binding, where for
 large values of $U$, the \uu contributions to coupling matrix elements $K^U_{cv,cv}$ in Eq.~\eqref{eq:e5} over-compensate for the sums of energy differences $\omega_{cv}$ and the Hartree+exchange-correlation contribution to  coupling matrix elements,  leading to 
unphysical complex eigenvalues. 
In Fig.~\ref{fig:nn-uuu-exc} 
we find that, when \u and \uu are combined consistently,  \uu primarily cancels the  effects of \u on $3d \rightarrow 3d$ type of excitations, which are in the $\sim$ 3.5 - 4.5 eV range.
This cancellation of \u by \uu gives rise to an approximately quadratic net dependence on $U$ within the full Casida equation,
as  opposed to a rather linear net behaviour with $U$ when the TDA is invoked. 
We can clearly observe this when comparing Fig.~\ref{fig:nn-uuu-exc} and TDA in Fig.~\ref{fig:nn-uuu-exc-tda}.
This, again, reflects what was previewed in our four-level toy model.
\begin{figure}[t]
\centering
 \includegraphics[width=0.2\textwidth]{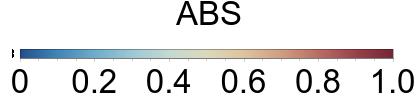} 
 
\subfloat[ \u\& TDDFT \label{fig:nn-uuo-abs}]{\includegraphics[width=0.22\textwidth]{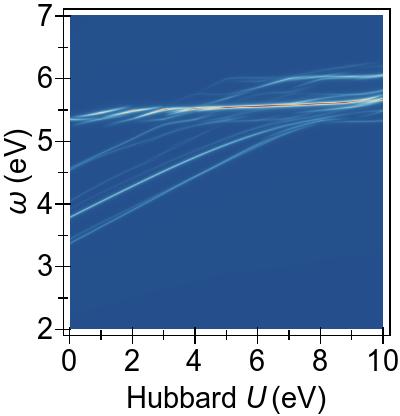}}\hfill
\subfloat[DFT  \& \uu \label{fig:nn-oou-abs}]{\includegraphics[width=0.22\textwidth]{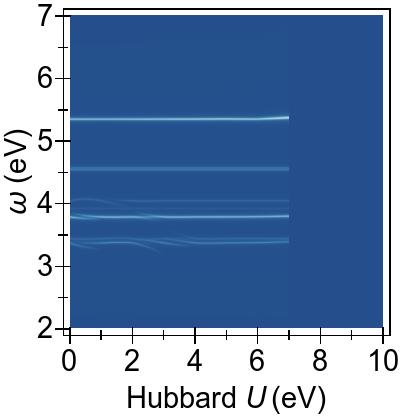}}

\subfloat[ \u\& \uu \label{fig:nn-uuu-abs}]{\includegraphics[width=0.22\textwidth]{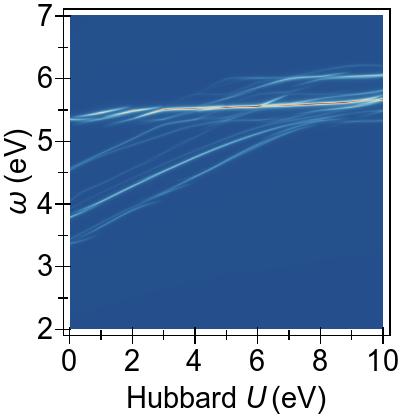}}\hfill
\subfloat[ \u \& \uu (TDA)  \label{fig:nn-uuu-abs-tda}]{\includegraphics[width=0.22\textwidth]{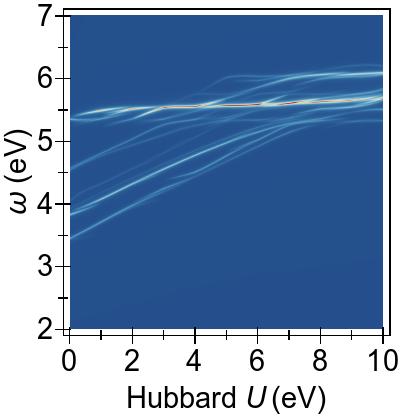}}

\caption{The dipole-dipole absorption spectra of  Ni(CN)$_4^{2-}$, calculcated using \u\& TDDFT, DFT \& TDDFT+$U$, and \u\& TDDFT+$U$, as  functions of the Hubbard $U$ parameter.}
\label{fig:nn-abs}
\end{figure}

Overall, on one hand \u is very efficient at modifying the ABS  as it pushes  low-lying optical transitions to higher energies, as seen in Fig.~\ref{fig:nn-uuo-abs}, Fig.~\ref{fig:nn-uuu-abs} and  Fig.~\ref{fig:nn-uuu-abs-tda}.
On the other hand, \uu does not have any significant effect 
at all on the ABS shown  in Fig.~\ref{fig:nn-oou-abs},  as \uu acts solely on $3d \rightarrow 3d$ excitations, which are  optically 
perfectly dark in  Ni(CN)$_4^{2-}$ here due to its idealized square-planar symmetry.

\subsection{The tetrahedral nickel tetracarbonyl: Ni(CO)$\mathbf{_4}$ }\label{sec:c8s3}
\begin{figure}[t]
\centering

\hfill \includegraphics[width=0.2\textwidth]{mol-ees-legend}

\subfloat[Energy levels \label{fig:no-band}]{\includegraphics[width=0.23\textwidth]{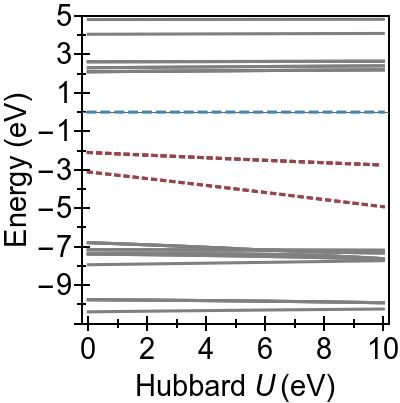}}\hfill
\subfloat[FGR \label{fig:no-fgr}]{\includegraphics[width=0.23\textwidth]{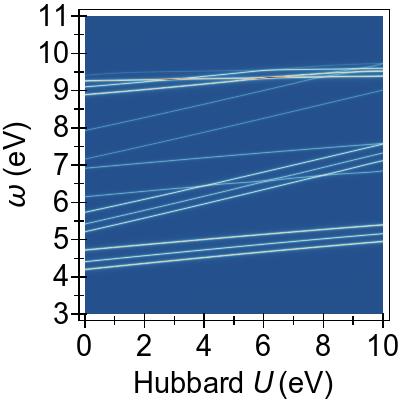}}

\caption{The Kohn-Sham DFT+$U$ energy levels and  singlet EES spectra 
of Ni(CO)$_4$ 
calculated using the FGR, as  functions of  Hubbard $U$ parameter. The Fermi level (dashed, blue line)  is set to 0 eV. The states most strongly 
affected by \u are shown with dashed, red lines.}
\label{fig:no-initial}
\end{figure}

The tetrahedral Ni(CO)\td{4} is  another low spin coordination with a neutral Ni center, 
but it is not perfectly isoelectronic with 
  Ni(CN)$_4^{2-}$
as it has an uncomplicated, full $3d$ sub-shell.
The (CO)$^-$ ion is a strong-field $\pi$-acceptor ligand, which splits the $3d$ states of Ni into $d_{z^2}\approx d_{x^2-y^2}<d_{xy}\approx d_{xz} \approx d_{yz}$ due to the tetrahedral symmetry present.
The two-fold and the three-fold degenerate $3d$ splitting can be clearly distinguished by the differing response to \u seen in Fig.~\ref{fig:no-band}. In this systems,  the  low-lying singlet excitations are necessarily of a predominantly Ni $3d \rightarrow \pi^*$ character~\cite{doi:10.1021/ja00202a004,doi:10.1021/acs.jpca.5b04844}.

In Fig.~\ref{fig:no-band}, we observe that the two-fold degenerate $d_{z^2}\approx d_{x^2-y^2}$ states (red, dashed line) at $-2$~eV and the three-fold degenerate   $d_{xy}\approx d_{xz} \approx d_{yz}$  states (red, dashed lines), at $-3$~eV for $U$= 0 eV, are pushed deeper with increasing $U$ values within DFT+$U$.  
In Fig.~\ref{fig:no-fgr}, these immediate effects of \u on the low-lying  $3d\rightarrow \pi^*$ excitations,
at $\sim 4.0 - 5.5$~eV for $U= 0$~eV,
 are reflected in  up-shifts in the 
FGR singlet EES with  increasing $U$ values. 
Such shifts are larger  for excitations from the $d_{z^2}$ and $d_{x^2-y^2}$ states, as these are lowered more by DFT+$U$.

\begin{figure}[t]
\centering
\includegraphics[width=0.2\textwidth]{mol-ees-legend}

\subfloat[\u \& TDDFT \label{fig:no-uuo-exc}]{\includegraphics[width=0.22\textwidth]{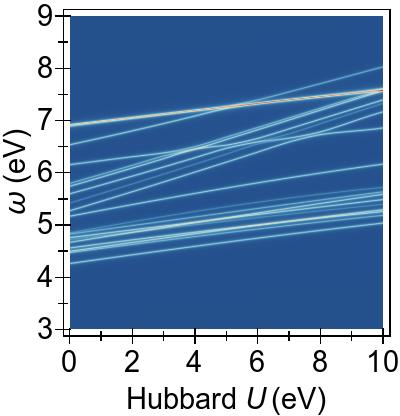}}\hfill
\subfloat[DFT  \& \uu \label{fig:no-oou-exc}]{\includegraphics[width=0.22\textwidth]{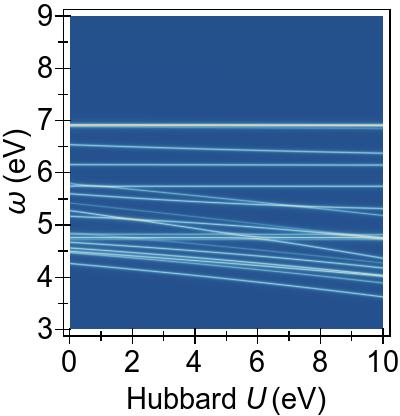}}

\subfloat[\u \& \uu \label{fig:no-uuu-exc}]{\includegraphics[width=0.22\textwidth]{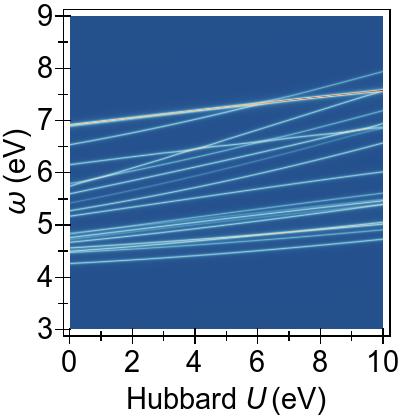}}\hfill
\subfloat[\u \& \uu (TDA) \label{fig:no-uuu-exc-tda}]{\includegraphics[width=0.22\textwidth]{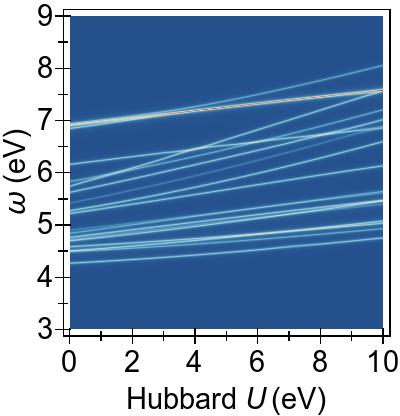}}

\caption{The singlet EES of Ni(CO)$_4$ calculated using \u\& TDDFT, DFT \& TDDFT$+U$, and \u\& TDDFT+$U$,  as  functions of the Hubbard $U$ parameter.}
\label{fig:no-exc}
\end{figure}

\begin{figure}[t]
\centering
 \includegraphics[width=0.2\textwidth]{mol-abs-legend} 

\subfloat[\u \& TDDFT \label{fig:no-uuo-abs}]{\includegraphics[width=0.22\textwidth]{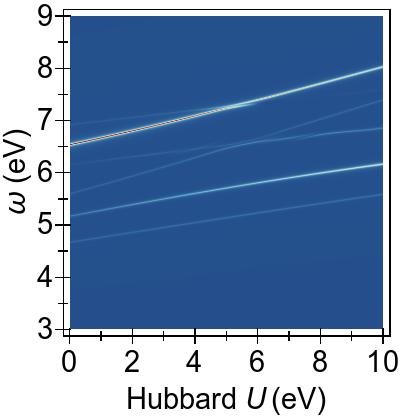}}\hfill
\subfloat[DFT  \& \uu \label{fig:no-oou-abs}]{\includegraphics[width=0.22\textwidth]{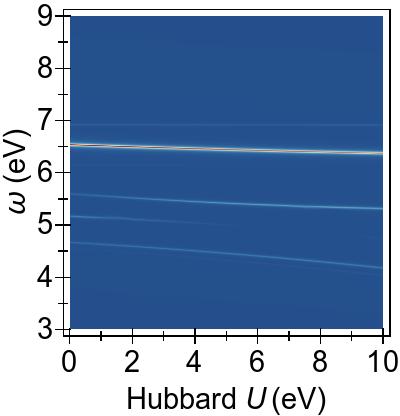}}

\subfloat[\u \& \uu \label{fig:no-uuu-abs}]{\includegraphics[width=0.22\textwidth]{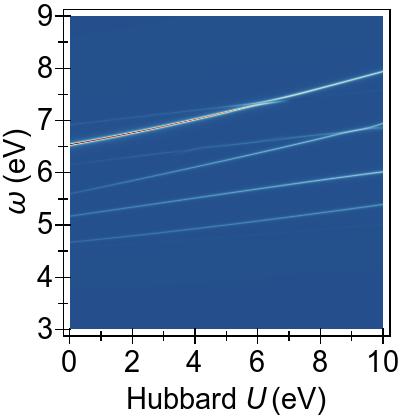}}\hfill
\subfloat[\u \& \uu (TDA) \label{fig:no-uuu-abs-tda}]{\includegraphics[width=0.22\textwidth]{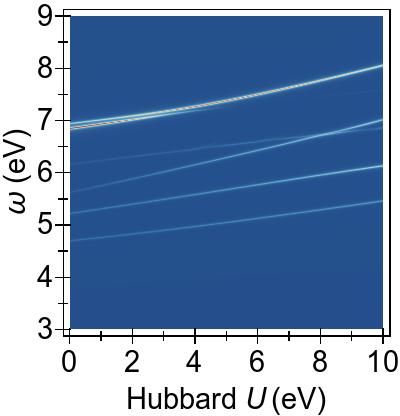}}

\caption{The dipole-dipole absorption spectra of  Ni(CO)$_4$ calculated using \u\& TDDFT, DFT \& TDDFT$+U$, and \u\& \uu   as  functions of the Hubbard $U$ parameter.}
\label{fig:no-abs}
\end{figure}

A complete picture of the behaviour of the first $50$ excitations with 
\u and \uu is presented in Fig.~\ref{fig:no-exc}. 
The increasing $U$ parameter in \u   affects excitation energies globally, by pushing them to higher energies. 
In Fig.~\ref{fig:no-uuo-exc}, particularly, the excitations from the lower-lying $3d$ levels ($d_{z^2}/d_{x^2-y^2} \rightarrow \pi^*$), at $\sim  5 - 6$~eV for $U$= 0 eV, climb most strongly and cross over with the excitations from  the deeper states at around $U \approx$ 4 eV,  as was previewed in Fig.~\ref{fig:no-fgr}.
A similar trend is also present with \u as it is more effective on the excitations from the lower energetic $3d$ levels, as seen in Fig.~\ref{fig:no-oou-exc}, where some  cross over   occurs with the lower-energy group of  excitations.
The cancellation of \u effects by \uu is more subtle in Ni(CO)$_4$ for the relevant excitations   compared to the situation in Ni(CN)$_4^{2-}$,  and this 
(shown in Fig.~\ref{fig:no-uuu-exc}) 
is as expected due to the weaker $3d\rightarrow 3d$ character of the transitions. 
While \uu shifts the lowest group of excitations  as well as splitting  these excitations, it does not lead to the splitting-off of distinct tightly-bound excitons as observed in  Ni(CN)$_4^{2-}$.  
As the   dominant optically allowed transitions are almost purely of $3d\rightarrow \pi^*$ character, \u naturally pushes  bright excitations up in energy, as seen in Fig.~\ref{fig:no-uuo-abs}, whereas  the effect of \uu on these excitations is quite subtle, which can be seen in Fig.~\ref{fig:no-oou-abs}.
An important point to recall here is that, while \u is effective in  proportion to the  $3d$ character of  the KS manifold, \uu is proportional to the $3d$ character of product space   of occupied $3d$-unoccupied $3d$ subspaces.

\section{First-principles spectra of two low-spin Nickel-centred complexes}
\label{section4}

The EES and ABS of our two closed-shell coordination complexes were generated using  \u and \uu with their respective  first-principles Hubbard $U_\textrm{eff}$ parameters, 
following the detailed procedure described in the Appendices.
In particular, these spectra were obtained 
by evaluating, or `slicing', the interpolated data shown in the graphs presented in 
Sec.~\ref{sec:c8s2} and Sec.~\ref{sec:c8s3}, at the corresponding 
first-principles Hubbard $U$ parameters summarised in Table~\ref{tab:c8t3}.

\subsection{Excitation energies and spectra of Ni(CN)$\mathbf{_4^{2-}}$}
\begin{figure}[t]
\centering
\includegraphics[width=0.45\textwidth]{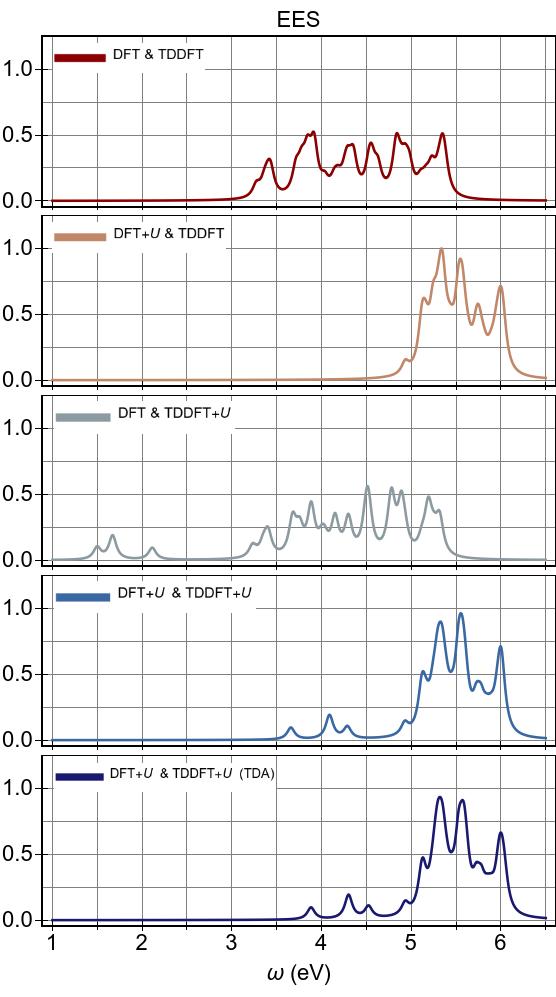}
\caption[The  singlet EES of Ni(CN)$_4^{2-}$  
calculated using a first-principles Hubbard $U$ parameter.]{The  singlet EES  of Ni(CN)$_4^{2-}$ extracted from Fig.~\ref{fig:nn-exc} by taking a cross-section at the first-principles Hubbard $U_\textrm{eff}=6.901$~eV, and shown with a Lorentzian broadening of $0.1$~eV. 
}
\label{fig:nn-exc-u}
\end{figure}

\begin{figure}[t]
\centering
\includegraphics[width=0.45\textwidth]{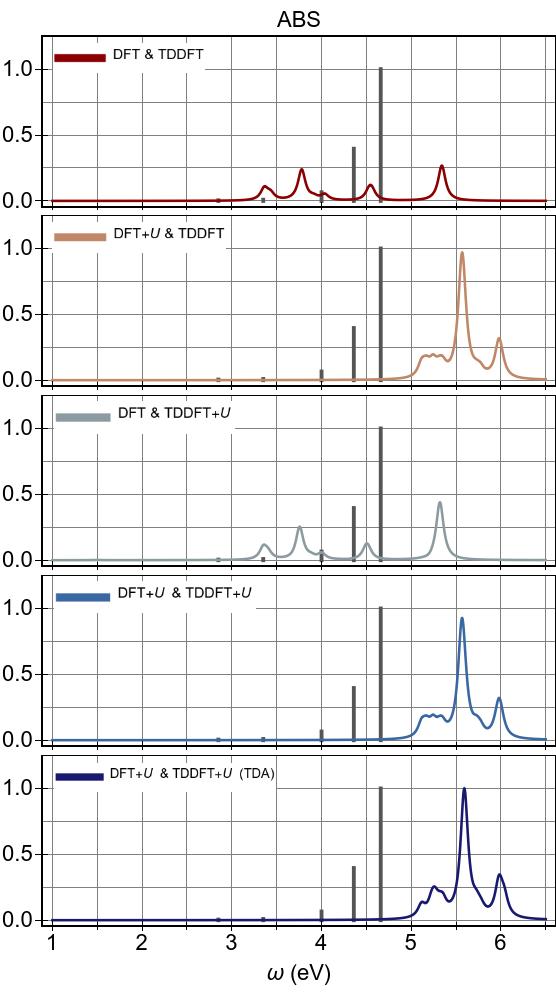}
\caption[The  singlet dipole-dipole ABS of Ni(CN)$_4^{2-}$ calculated using a first-principles Hubbard $U$ parameter.]{The  singlet dipole-dipole ABS of Ni(CN)$_4^{2-}$  extracted from Fig.~\ref{fig:nn-abs} by taking a cross-section at the first-principles Hubbard $U_\textrm{eff}=6.901$ eV, and shown with a Lorentzian broadening of $0.1$~eV.  The experimental absorption energies extracted from Ref.~\citenum{doi:10.1021/ja00886a002} are shown with vertical grey lines that are scaled with respect to the maximum absorbance of the highest energy peak at $4.66$~eV.}
\label{fig:nn-abs-u}
\end{figure}

The EES and ABS of Ni(CN)$_4^{2-}$ are presented in Fig.~\ref{fig:nn-exc-u} and Fig.~\ref{fig:nn-abs-u}  for the first-principles $U_\textrm{eff}= U - J = 6.901$~eV, alongside  experimental excitation spectra  extracted from Ref.~\citenum{doi:10.1021/ja00886a002}.  
In Fig~\ref{fig:nn-abs-u}, 
the experimental excitation peak positions are shown with vertical grey lines, 
with heights indicating their relative absorbances with respect to that of 
 the experimental maximum absorbance at $4.66$~eV, which is set to unity.
Excitation energies are listed in Table~\ref{tab:c8t5}  along with the experimental
 results~\cite{doi:10.1021/ja00886a002} and TDDFT results~\cite{doi:10.1021/ic060584r}, 
with  optically bright excitations are highlighted with a  bold font.
In particular, our
first-principles excitation energies were obtained from the peak positions of Fig.~\ref{fig:nn-exc-u}, with smaller  peaks and shoulders removed, and the optically bright ones were assigned by matching to the peaks of Fig.~\ref{fig:nn-abs-u}.
The previous TDDFT calculations of Ref.~\citenum{doi:10.1021/ic060584r} were performed using implicit solvation with a dielectric constant of $37.5$, 
whereas ours were performed under vacuum conditions.
Nonetheless, the former data provides an useful 
benchmark for testing  the numerical validity of our \uu code.
As seen Fig.~\ref{fig:nn-exc-u}, \u is effective throughout the spectral range. 
It  shifts excitation features to higher energies, as seen by comparing \u \& TDDFT with DFT \& TDDFT (PBE).  
TDDFT+$U$, however, acts only in the low-energy range, 
and it gives rise to the emergence of new peaks surrounded  by those already present in DFT \& TDDFT. 
The combined effects of \u and \uu proves to be almost a simple combination of their respective individual effects,  as seen in EES with \u \& TDDFT+$U$, where  excitation energies are globally shifted and some additional peaks emerge.

In Fig.~\ref{fig:nn-abs-u} (also represented  in Table~\ref{tab:c8t5}), regardless of its flavour, TDDFT fails to capture the optically bright excitation at $2.85$~eV observed in experiment, and this is consistent 
with previous TDDFT studies using the LDA and PBE functionals.
Hybrid TDDFT using the B3LYP functional performs  better than LDA  or PBE
in this regard, surely due to its better (more spatially long-ranged) 
description of exciton binding via its partial inclusion of the exact exchange interaction.
In  Fig.~\ref{fig:nn-abs-u}, we see that \u  carries optically bright features to higher energies and dramatically changes the overall appearance of the spectrum.
In fact, \u clearly worsens the agreement with 
experimental excitation energies, 
by pushing excitations within DFT \& TDDFT to 
higher energies such that the lowest optically bright excitation is carried to a position $\sim 1.8$~eV higher energy compared to 
that of DFT \& TDDFT.
We find that \uu has a relatively minor effect on the optically bright excitations
when applied upon DFT (PBE), and no discernible effect when applied
upon DFT+$U$.
Thus, \uu does not mitigate the harmful effects of \u  on optically
bright excitations in this system.
TDA and RPA predict  spectra in close mutual agreement,  with slightly higher energies emerging within TDA for both spectra.

\begin{table}[H]
\renewcommand{\arraystretch}{1.2} \setlength{\tabcolsep}{4pt}
\begin{center}
{\scriptsize
\begin{tabular}{lclclclclclclclc|}
\hline \hline
Method  & & &  & & &
  \\ 
\hline 

\splitbox{\\ DFT \& \\ TDDFT (PBE)}    & \splitbox{\\ \bf{3.37} \\ 4.34}  & \splitbox{\\ 3.42 \\ \bf{4.55}} & \splitbox{\\ \bf{3.78}\\\ 4.84} & \splitbox{\\ 3.85 \\ 4.92} & \splitbox{\\ 3.91 \\ 5.23 }& \splitbox{\\ \bf{4.03}\\ \bf{5.34}} \\ \\  \hline

\splitbox{\\ \u \& \\ TDDFT } & \splitbox{\\ 4.94 \\ \bf{5.98} }  & \splitbox{\\ \bf{5.17} \\   } & \splitbox{\\ \bf{5.24} \\   } & \splitbox{\\ \bf{5.33} \\   } & \splitbox{\\ \bf{5.57}\\  } & \splitbox{\\ 5.74 \\  } \\ \\  \hline

\splitbox{\\ DFT \& \\ \uu }   & \splitbox{\\ 1.50 \\  \bf{3.76} \\  \bf{4.51}}  & \splitbox{\\ 1.67 \\ 3.88  \\ 4.78  } & \splitbox{\\ 2.11 \\ \bf{4.00}  \\ 4.89  } & \splitbox{\\ 3.24 \\ 4.15  \\ 5.19 } & \splitbox{\\ \bf{3.36} \\ 4.30  \\  \bf{5.32}} & \splitbox{\\ 3.68 \\ \\ } \\ \\ \hline

\splitbox{\\ \u  \& \\ \uu }  & \splitbox{\\ 3.66 \\ \bf{5.33} }  & \splitbox{\\ 4.09 \\ \bf{5.57} } & \splitbox{\\ 4.29 \\ 5.75  } & \splitbox{\\ 4.93 \\  \bf{5.98} } & \splitbox{\\ \bf{5.17} \\ } & \splitbox{\\ \bf{5.24} \\  } \\ \\  \hline

\splitbox{\\ \u \& \\ \uu (TDA)}  & \splitbox{\\ 3.88 \\ \bf{5.59} }  & \splitbox{\\ 4.30 \\ 5.73 } & \splitbox{\\ 4.52 \\  \bf{5.98} } & \splitbox{\\ 4.94 \\ } & \splitbox{\\ \bf{5.12} \\ } & \splitbox{\\ \bf{5.26} \\ }

\\
 & & &  & & & \\
\hline 
&  & &  & & & \\
Exp.~\cite{doi:10.1021/ja00886a002} &
 \bf{2.85} & \bf{3.35} & \bf{4.00} & \bf{4.36} & 
 \bf{4.66} & \\ \\
 TDDFT (PBE)~\cite{doi:10.1021/ic060584r} &   \bf{3.99} & 
\bf{4.19} & \bf{4.48}  & \bf{3.76} & \bf{4.12} & 4.53 \\ \\
TDDFT (LDA)~\cite{doi:10.1021/ic060584r}  &  \bf{3.98} & \bf{4.17} & 
\bf{4.46} & \bf{3.78} & \bf{4.13} & 4.55 \\ \\
TDDFT (B3LYP)~\cite{doi:10.1021/ic060584r}  &  \bf{3.29} & \bf{3.57} & \bf{3.92} & \bf{4.75} & \bf{5.07} & 5.59 \\ 
& &  &  & & & \\
 \hline \hline
\end{tabular}}
\end{center}
\caption[Energies (in eV) of the singlet excitations of Ni(CN)$_4^{2-}$.]{Energies (in eV) of the singlet excitations of
Ni(CN)$_4^{2-}$, as obtained  
without symmetry 
assignment from 
the peak positions of Fig.~\ref{fig:nn-exc-u}, with smaller  peaks and shoulders removed.
Coinciding peaks in Fig.~\ref{fig:nn-abs-u}   are assigned as optically bright excitations and highlighted with a bold font. }
\label{tab:c8t5}
\end{table}

\subsection{Excitation energies and spectra of  Ni(CO)$\mathbf{_4}$}
\begin{figure}[t]
\centering
\includegraphics[width=0.45\textwidth]{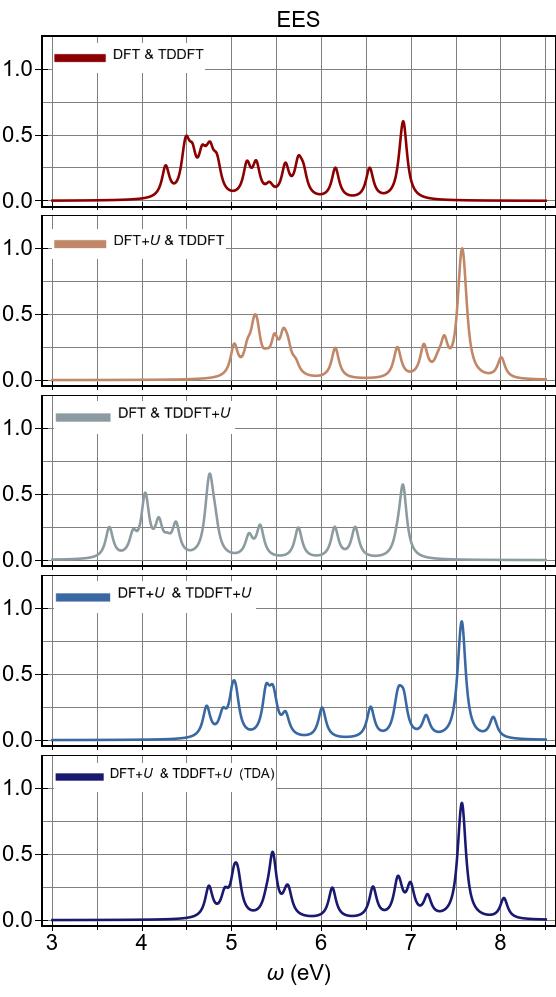}
\caption[The  singlet EES  of Ni(CO)$_4$ calculated using a first-principles Hubbard $U$ parameter.]{The  singlet EES  of Ni(CO)$_4$ extracted from Fig.~\ref{fig:no-exc} by taking a cross-section at the first-principles Hubbard $U_\textrm{eff}=9.849$~eV, 
and shown with a Lorentzian broadening of $0.1$ eV. The  EES (grey, dashed line), constructed from INDO/S quantum-chemical excitation energies extracted from Ref.~\citenum{doi:10.1021/ja00202a004}, is shown with a Lorentzian broadening of $0.1$~eV.}
\label{fig:no-exc-u}
\end{figure}

\begin{figure}[t]
\centering
\includegraphics[width=0.45\textwidth]{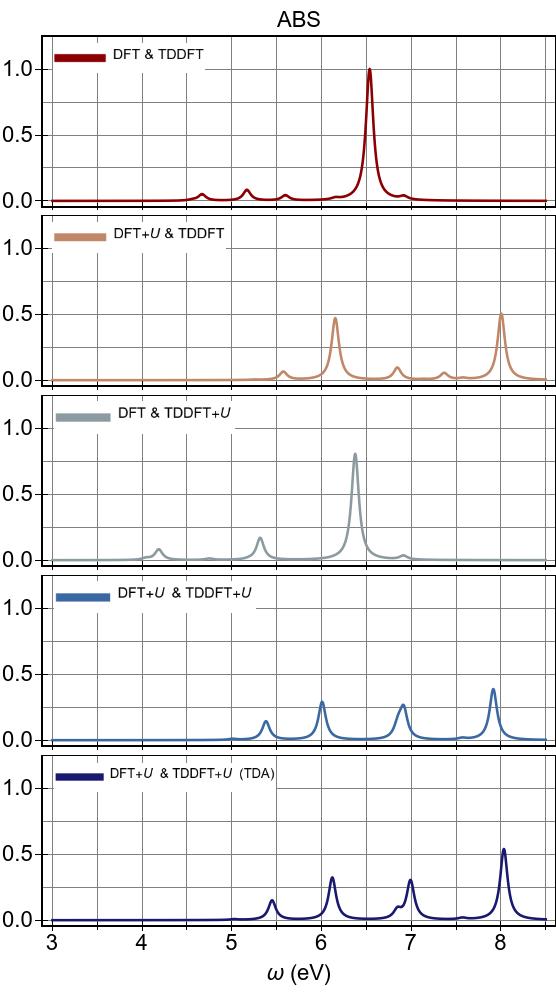}
\caption{The  singlet dipole-dipole ABS of Ni(CO)$_4$ extracted from Fig.~\ref{fig:no-abs} by taking a cross-section at the first-principles Hubbard $U_\textrm{eff}=9.849$ eV, and shown with a Lorentzian broadening of $0.1$ eV. The  ABS (grey, dashed line), constructed from INDO/S quantum-chemical excitation energies and oscillator strengths extracted from Ref.~\citenum{doi:10.1021/ja00202a004}, is shown with a Lorentzian broadening of $0.1$~eV.}
\label{fig:no-abs-u}
\end{figure}

The EES and ABS of Ni(CO)$_4$ are presented in Fig.~\ref{fig:no-exc-u} and Fig.~\ref{fig:no-abs-u}, respectively,  for the first-principles $U_\textrm{eff}=9.849$~eV. Shown alongside, for comparison, are  the corresponding spectra generated 
using the experimental excitation energies and oscillator strengths  extracted from Ref.~\citenum{doi:10.1021/ja00202a004}.
In this molecule, due to its less-than-full $3d$ manifold
and hence increased $3d$ character of the valence-conduction 
transition space,
we will see that \uu is rather more effective than it is in the case of  
Ni(CN)$_4^{2-}$. 
However, it is still not enough to compensate for the inaccuracy that the 
contemporary DFT+$U$ potential introduces and, intriguingly, 
DFT \& TDDFT+$U$ performs by far the best among the combinations
tested.

In Fig.~\ref{fig:no-abs-u}
(also in Table~\ref{tab:c8t6}, we observe that DFT \& TDDFT overestimates the lowest optically bright excitation by $\sim 1.1$~eV compared to in-vacuo 
INDO/S
(the intermediate neglect of differential overlap model adapted
for spectroscopy) quantum-chemical calculations.
DFT+$U$ worsens this over-estimation to 
$\sim 1.4$~eV, while arguably also worsening the line-shape agreement.
TDDFT+$U$ applied upon this (DFT+$U$ \& TDDFT+$U$)
 makes relatively little difference,
 and the effect of invoking the TDA is approximately that of  a small, 
rigid  blue-shift.
It is difficult to make
a clear comparison against the large spread of experimental values, meanwhile.
 The  agreement between the peak positions and line-shapes
 (we do not attempt to compare physical magnitudes here)
 given by DFT \& TDDFT+$U$ and INDO/S, both for EES and ABS, is remarkable,
 however,  
 with the first bright energy agreeing to $\sim 0.04$~eV
 (albeit with a splitting in INDO/S that is absent in TDDFT+$U$).
The ABS peak positions are also in reasonable agreement with some 
of the experimental values given in Table~\ref{tab:c8t6}, though again interpretation is challenging
here due to the spread of values.
We now digress to consider these results.

\subsection{The use of a single Hubbard $U$ parameter
in DFT+$U$ and TDDFT+$U$}
\label{mycaveat}

The improvement of DFT \& TDDFT by a first-principles Hubbard $U$
correction to the \emph{kernel} but not to the \emph{potential},
 if INDO/S can be taken
as a benchmark, may be understood as a possible consequence of the 
following.
The Hubbard $U$ parameter is a measure of spurious interaction, 
one that is calculated as the derivative of an averaged potential which, in turn, 
is  a measure of the derivative of an energy. 
On one hand, therefore, $U$ is well 
suited to measure the magnitude required for correction of
the interaction kernel.
On the other hand, it is not necessarily a good measure of the magnitude
required for correction of the Kohn-Sham potential.
More specifically, it has recently been shown by one of the present authors 
that very different parameters $U_1$ and $U_2$ may be needed for the constant
and linear terms in the density, respectively, of the DFT+$U$ corrective potential~\cite{PhysRevB.94.220104}.
Put another way, the linear and quadratic terms in Eq.~\ref{eq:e3}
may benefit from different $U_\textrm{eff}$ pre-factors.

Dubbed DFT+$U_1$+$U_2$, this generalization of \u allows  for 
the approximate enforcement of Koopmans' condition on the DFT+$U$
subspace, which is a condition that is 
implied by the assumptions 
under-pinning the calculation of $U$.
In other words, while the Hubbard $U$ may successfully measure the
self-interaction strength, and possibly open the correct fundamental gap via the quadratic energy term, 
a single parameter does not 
carry enough information to correctly position the targeted subspace energetically with respect to the background (also known as bystander) states, a task for which the linear
term is better equipped.
Put yet another way, the double-counting 
correction used in the derivation 
of the contemporary DFT+$U$ functional is arguably too simple, 
for certain system types, and could gainfully 
by given its own separate pre-multiplicative parameter.
The TDDFT+$U$ kernel does not suffer from this complication,
however, since
only the usual parameter associated 
with the quadratic energy term survives in the kernel. 
In this sense, contemporary methods for 
calculating a single $U$ 
parameter may actually be better suited to 
TDDFT+$U$ than to DFT+$U$.
This is reflected by the
apparently, paradoxically superior performance
of DFT \& TDDFT+$U$ over DFT+$U$ \& TDDFT+$U$
in the aforementioned system Ni(CO)$_4$, albeit that this is
a rather extreme test of DFT+$U$ insofar as that 
the uncorrected PBE functional already performs well, 
and that the relevant subspace is very far from half-filling.

Indeed, any ill-effects of conventional \u on 
the potential are expected to be most 
strongly felt when applying DFT+$U$ to spin-unpolarized $3d$ 
spaces that are almost full (or empty) such as in 
Ni(CO)$_4$, since then the  conduction (or valence) 
band edge is of predominantly background-orbital
character. 
The Kohn-Sham gap is neither of $3d \rightarrow 3d$ character
nor reliably determined by the familiar $U$ in such cases.
A work-around alternative (albeit not equivalent) to DFT+$U_1$+$U_2$
may be the application of DFT+$U$ to other orbital types, e.g.
O $2p$, C $2p$, and possibly Ni $4s$, but this has not been explored in the
present work.
A complete counter-example to this, where \u is very effective, is next provided by an 
open-shell  complex, where the Kohn-Sham gap is 
strongly affected by a 
varying Hubbard $U$ parameter.

\begin{table}[H]
\renewcommand{\arraystretch}{1.0} \setlength{\tabcolsep}{4pt}
\begin{center}
{\scriptsize
\begin{tabular}{lclclclclclclclc|}
\hline \hline
Method & & &  & & & \\ \hline
\splitbox{\\ DFT \& \\ TDDFT (PBE)}  & \splitbox{\\4.26 \\ 5.42 }  & \splitbox{\\4.50 \\ \bf{5.60}} & \splitbox{\\ \bf{4.67}\\ 5.75} & \splitbox{\\ 4.75 \\\bf{6.16}} & \splitbox{\\ \bf{5.17}\\\bf{6.54}} & \splitbox{\\5.27\\ \bf{6.91}} \\   \\  \hline

\splitbox{\\ \u \& \\ TDDFT } & \splitbox{\\5.03 \\ 7.14 }  & \splitbox{\\5.26 \\ \bf{7.37}} & \splitbox{\\ 5.48 \\ 7.57 } & \splitbox{\\ \bf{5.57} \\ \bf{8.00} } & \splitbox{\\ \bf{6.15} \\ } & \splitbox{\\ \bf{6.85} \\  } \\   \\ \hline

\splitbox{\\ DFT \& \\ \uu } & \splitbox{\\3.63 \\ 5.19}  & \splitbox{\\ 3.91 \\ \bf{5.32} } & \splitbox{\\ 4.03 \\ 5.74} & \splitbox{\\ \bf{4.19} \\ 6.15} & \splitbox{\\ 4.38 \\ \bf{6.38} } & \splitbox{\\ 4.75 \\ \bf{6.91} } \\  \\  \hline

\splitbox{\\ \u \& \\ \uu } & \splitbox{\\ 4.72 \\ \bf{6.01} }  & \splitbox{\\ 4.91 \\ 6.55 } & \splitbox{\\ 5.02 \\ \bf{6.91} } & \splitbox{\\ \bf{5.39} \\ 7.17 } & \splitbox{\\ 5.45 \\ 7.57 } & \splitbox{\\ 5.59 \\ \bf{7.92} } \\   \\  \hline

\splitbox{\\ \u \& \\ \uu (TDA) } & \splitbox{\\ 4.75 \\ 6.57 }  & \splitbox{\\ 4.93 \\ \bf{6.86} } & \splitbox{\\ 5.04 \\ \bf{6.99}} & \splitbox{\\ \bf{5.45} \\ 7.18 } & \splitbox{\\ 5.62 \\ 7.56} & \splitbox{\\ \bf{6.12}\\ \bf{8.03}  } \\ 
& & & &  & & \\
\hline
& & &   & & & \\
Exp. (solvent)~\cite{doi:10.1021/ja01015a013}  &  & \bf{5.24} & \bf{5.52} & \bf{6.02} &  &   \\  \\
Exp. (matrix)~\cite{doi:10.1021/ic50198a007}  & \bf{4.54} & \bf{5.17} & 
&  &  &   \\ \\
Exp. (gas)~\cite{doi:10.1021/ja00202a004}  & \bf{4.5} &  &  \bf{5.4} &   \bf{6.0} & 
&  \\  \\  \hline
\splitbox{\\ INDO/S ~\cite{doi:10.1021/ja00202a004} \\} & \splitbox{\\ 3.93 \\  4.55 \\  5.29}  & \splitbox{\\ 3.98 \\ 4.64  \\ \bf{5.36} } & \splitbox{\\ 4.05 \\ 4.79  \\ 5.71  } & \splitbox{\\ \bf{4.15} \\ 4.91  \\ \bf{6.20} } & \splitbox{\\ \bf{4.36} \\ 4.95 \\  } & \splitbox{\\ 4.39 \\ 5.11 \\ } \\ \\ \hline
TDDFT (LDA)~\cite{doi:10.1021/jp991060y}  & \splitbox{\\ 4.36 \\ \bf{5.37}}  & \splitbox{\\ 4.60 \\ \bf{5.84}} & \splitbox{\\ \bf{4.62} \\ \bf{6.01}} & \splitbox{\\ \bf{4.70} \\ } & \splitbox{\\ \bf{4.82} \\ } & \splitbox{\\ 4.95\\  } \\  \\ \hline
 SAC-CI~\cite{doi:10.1063/1.470325}  & \splitbox{\\ 4.52 \\ \bf{5.51}}  & \splitbox{\\ 4.53 \\ \bf{5.72}} & \splitbox{\\ \bf{4.79} \\ \bf{5.76}} & \splitbox{\\ 4.97 \\ 6.07} & \splitbox{\\ 5.25 \\ 6.28} & \splitbox{\\ 5.41 \\ } \\  \\ \hline
CASPT2~\cite{doi:10.1021/jp9614355} & \splitbox{\\ 3.58 \\ 5.15}  & \splitbox{\\ 3.72 \\ 5.20} & \splitbox{\\ 4.04 \\ \bf{5.22}} & \splitbox{\\ \bf{4.34} \\ \bf{5.57}} & \splitbox{\\ 4.88 \\ 6.00} & \splitbox{\\ 5.14 \\ 6.01 } \\ 
 & & & & & & \\ 
\hline \hline
  & & & & &  & \\
\end{tabular}}
\end{center} 
\caption[Energies (in eV) of the singlet excitations of
Ni(CO)$_4$.]{Energies (in eV) of the singlet excitations of
Ni(CO)$_4$, as obtained  without symmetry assignment from 
the peak positions of Fig.~\ref{fig:no-exc-u}, with smaller  peaks and shoulders removed.
The coinciding peaks in Fig.~\ref{fig:no-abs-u}  are  assigned as optically bright excitations highlighted with a bold font. }
\label{tab:c8t6}
\end{table}

\section{First-principles spectra of a high-spin Cobalt-centred complex}
\label{Section5}

\begin{figure}[H]
\centering
\tcbox[sharp corners, boxsep=0.0mm, boxrule=0.5mm,  colframe=gray, colback=white]{\includegraphics[width=.25\textwidth]{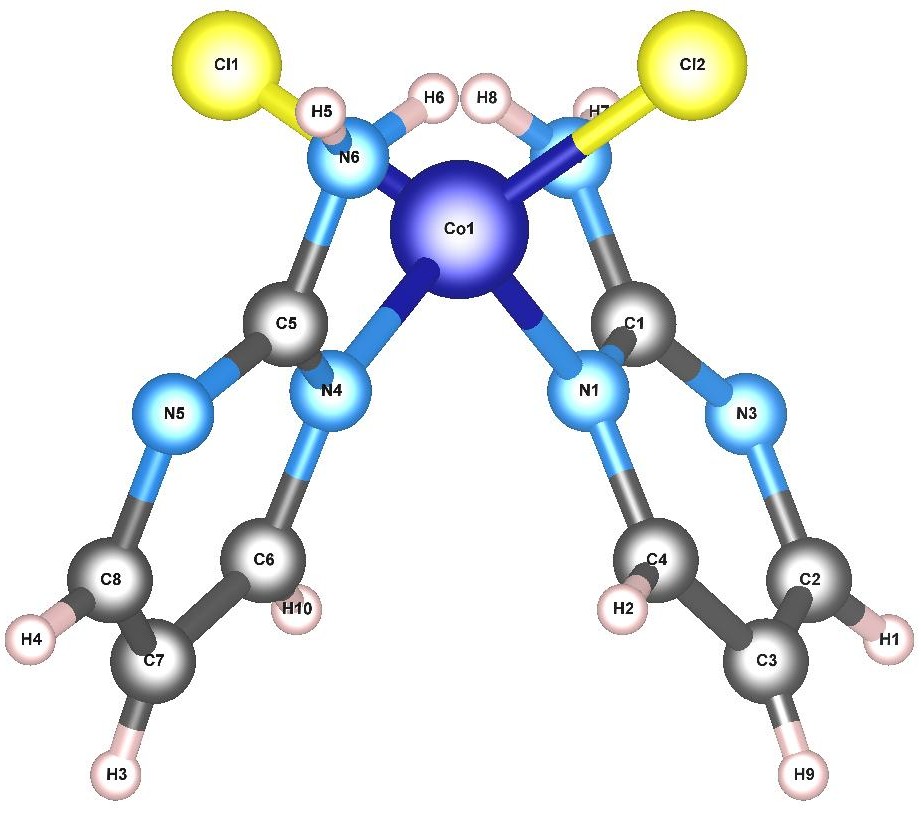}}

\caption{The molecular structure of CoL$_2$Cl$_2$ (L=2-aminopyrimidine: C$_4$H$_5$N$_3$).}
\label{fig:co-str}
\end{figure}

CoL$_2$Cl$_2$ (L=2-aminopyrimidine: C$_4$H$_5$N$_3$) is a Co-centred,
distorted pseudo-tetrahedral complex with  two types of ligands, as  illustrated in Fig.~\ref{fig:co-str}. 
The central Co atom has a nominal charge of  $2+$,  with a $3d$ sub-shell 
containing  $7$ electrons.
Cl$^{-}$ is a $\pi$-donor weak-field ligand, which leads to a splitting of the $3d$ sub-shell of the  Co atom into a high-spin configuration in a pseudo-tetrahedral symmetry~\cite{griffith1964theory,B902743B}.
In its  high-spin configuration, the $3d$ orbitals at higher energies
contain unpaired electrons, resulting  in a total 
spin of $3~\mu_\textrm{B}$.
The fully and partially filled molecular orbitals at higher energies are predominantly hybrids comprised of Co $3d$  and    Cl $3p$ orbitals. 
Moreover, further splitting in the  energy levels by $3d-3p$ hybridisation occurs by means of the distortion due to the tilted L-ligands. 
The low-lying excitations are expected to have strong $3d\rightarrow 3d$ character in this molecule.

Experimental values for the low-lying, spin-allowed optically bright excitations of CoL$_2$Cl$_2$ are presented in Table~\ref{tab:c8t8}. 
Also provided are prior predictions from high-level quantum-chemistry methods, i.e., complete active space self-consistent field (CASSCF)
and CASSCF improved further by second-order N-electron valence perturbation theory (NEVPT2)~\cite{
doi:10.1021/ic200196k,doi:10.1063/1.1778711,doi:10.1063/1.1361246,doi:10.1063/1.1515317},  which were extracted from Ref.~\citenum{doi:10.1021/ic400980b}. 
Our own TDDFT calculations invoke the TDA for this spin-polarized system, 
due to technical limitations of the implementation.
Two different first-principles effective parameters 
were tested in our \u and \uu calculations, and
these were generated following the procedures
described in detail in Ref.~\citenum{linscott2018role}. 
Briefly, the like-spin $U_\textrm{eff} = U - J$
results from a treatment of the spin channels as forming an 
effective 2-site model in the `scaled $2 \times 2$' method,
and this is expected to yield results 
(in this case $5.724$~eV) comparable to those from
any correct method that separately calculates the Hubbard $U$
and Hund's $J$.
The less canonical `averaged $1 \times 1$' method
calculates the like-spin $U_\textrm{eff}$
as the average of the $U$ parameters calculated individually for the two spin channels when decoupled (each forming part of the bath for the other), 
and it may be a more reasonable 
assumption when an explicit $J$ correction term
is not used (as in the present work, where $U_\textrm{eff} = 3.798$~eV).

\begin{figure}[H]
\centering
\includegraphics[width=0.4\textwidth]{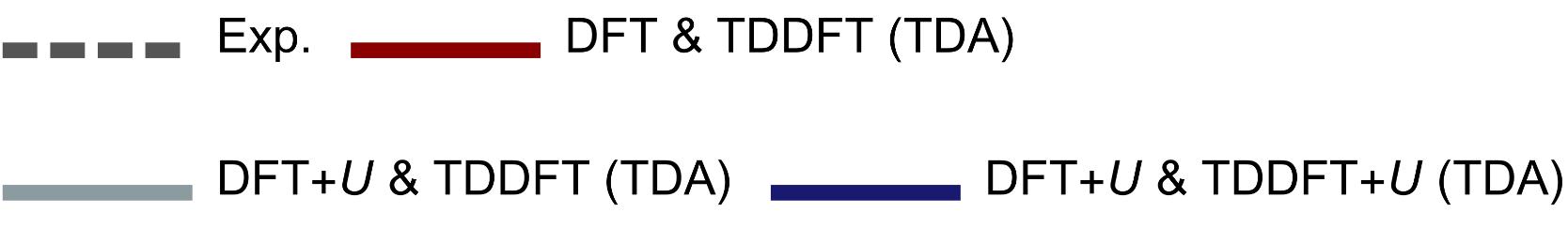} 
\vspace{-0.2cm}

\subfloat[\label{fig:co-exc}]{\includegraphics[width=0.45\textwidth]{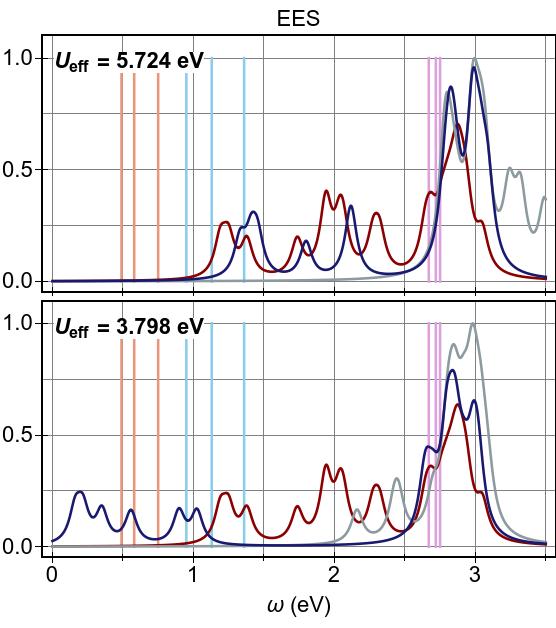}}

\subfloat[\label{fig:co-abs}]{\includegraphics[width=0.45\textwidth]{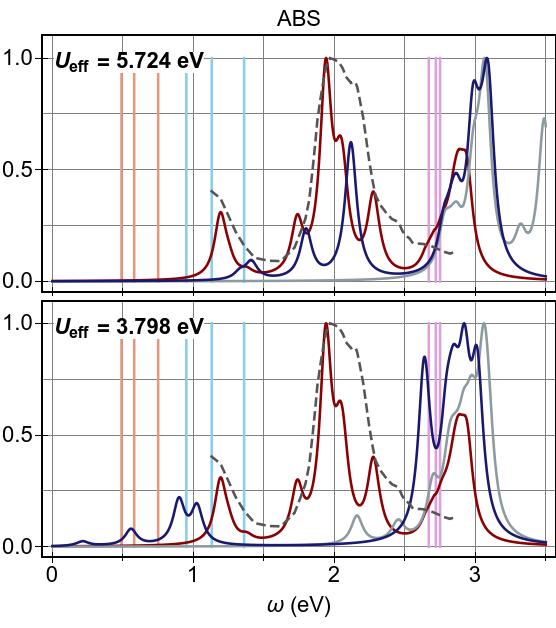}}

\caption{The  singlet EES and ABS of  CoL$_2$Cl$_2$ calculated using first-principles like-spin Hubbard $U_\textrm{eff}$ values calculated as described in
Appendix~\ref{sec:c8s4}, shown with a Lorentzian broadening of $0.1$ eV. 
The experimental absorption spectrum (grey, dashed line) was extracted from Ref.~\citenum{doi:10.1021/ic400980b} and scaled by setting the global peak to unity. The optically bright excitation energies calculated using CASSCF+NEWPT2 in Ref.~\citenum{doi:10.1021/ic400980b} are shown in the form of vertical lines with colors matching their values presented in Table.~\ref{tab:c8t8}.}
\label{fig:co-abs-exc}
\end{figure}

In Fig.~\ref{fig:co-exc}, we see that  \u \& TDDFT pushes 
excitation features at lower energies higher, compared to DFT \& TDDFT,
by   $\sim 1.6 - 2.0$~eV ($\sim 1.0$~eV) in the $2 \times 2$ 
($1 \times 1$) case.
In both cases an aggregate of
excitations forms at $\sim 2.8$~eV, and in neither
case does  \u \& TDDFT provide a promising agreement with
prior experiment or CASSCF-based results.
Meanwhile, the alternate combination, DFT \& TDDFT+$U$,
which performed rather well in the case of Ni(CO)$_4$, 
was found to be
not at all viable here, for either $U_\textrm{eff}$ value, as it gives rise
to unphysical, negative-valued excitation energies 
(a single instability).
The interaction of  \u \& \uu in this system is non-trivial, and 
the net result cannot
be well described as a linear combination 
(a cancellation) of the two method's effects, 
in general.
The linear combination picture holds to a greater degree for the 
\emph{higher}-valued, more canonical ($2 \times 2$) prescription for $U_\textrm{eff}$, counter-intuitively.
With this, we find that uncorrected DFT \& TDDFT does a better
job of reproducing the experimental absorption curve in 
Fig.~\ref{fig:co-abs}, and that the absent low-lying, tightly-bound exciton 
features predicted by CASSCF are no better recovered.
Here, referring to Fig.~\ref{fig:co-abs}, we emphasise that all curves
are independently normalised so that their maximum peak reaches a 
value of unity, and that it is not necessarily the case that 
DFT \& TDDFT recovers the experimental maximum absorption cross-section by any means.

Conversely, with the lower-valued, 
($1 \times 1$) prescription for $U_\textrm{eff}$, we find that the linear
combination picture breaks down completely.
With this  $U_\textrm{eff}$, it appears that the effect of
DFT+$U$ is insufficient to eradicate the strong 
$3d \rightarrow 3d$ character of the low-lying excitations.
Then, when \uu is applied on top of this, a very strong exciton
re-binding effect (of $\sim 2.0$~eV) occurs, yielding a net
exciton binding effect of  $\sim 1.0$~eV with respect to DFT \& TDDFT.
Ultimately,   \u \& \uu within the $1 \times 1$ prescription for
$U_\textrm{eff}$ does yield a group of tightly-bound ligand-field 
excitations that can be said to be in qualitative agreement with the
CASSCF predictions of Ref.~\citenum{doi:10.1021/ic400980b}.
The accuracy improvements for lower-energy excitations offered by \u \& \uu are  seen in Table~\ref{tab:c8t8}. 
Specifically, both  DFT \& TDDFT and \u \& TDDFT fail to capture the lowest three-fold degenerate excitation (highlighted with light pink) between $\sim 0.50 - 0.75$~eV predicted at the level of CASSCF+NEWPT2).
Moreover, \u \& TDDFT also overestimates the second group of three-fold degenerate excitations (highlighted with light blue) at around $\sim 0.90-1.40$ eV,  either when compared against  the experimental value of  $1.12$ eV or the CASSCF+NEWPT2 prediction of  $\sim 0.95 - 1.13$ eV.
\u \& \uu determines the lowest optically bright excitation energy with a relatively high accuracy at $0.56$ eV, comparing to both CASSCF and  CASSCF+NEWPT2.
Furthermore, it performs well by locating the second group of three-fold degenerate excitations (highlighted with light blue) at $0.90$ eV and $1.02$ eV.
%
However, \u \& TDDFT performs better, without a doubt, for the third 
group of three-fold degenerate excitations (highlighted with light purple) at 
$2.45$ eV, when comparing to the experimental value.
%
Overall, we can conclude that first-principles ($1 \times 1$ prescription)
\u\& \uu performs  better for low-lying excitations than DFT \& TDDFT, 
but this comes at the expense of completely removing the 
prominent absorption peak at $\sim 2.0$~eV where experiment and 
DFT \& TDDFT agree.
None of the available methods (including CASSCF), 
therefore, offer reliable correction of the spectra 
for both $3d \rightarrow 3d$ and 
higher-energy excitations, and this
is as expected given the spatially localized nature of Hubbard $U$
corrections when applied to metal $3d$ orbitals only.

\begin{table}[t]
\renewcommand{\arraystretch}{1.3} \setlength{\tabcolsep}{4pt}
\begin{center}
{\scriptsize
\begin{tabular}{lclclclclclc}
\hline \hline
Method & & &  & & & \\ \hline
\splitbox{\\ DFT \& \\ TDDFT (TDA) }  & \splitbox{\\ \bf{1.19} \\ \bf{2.28}}  & \splitbox{\\ 1.38 \\ 2.68} & \splitbox{\\ \bf{1.74} \\ \bf{2.92}} & \splitbox{\\ \bf{1.94} \\ } & \splitbox{\\ \bf{2.04} \\ } \\  \\ \hline

& & &  & & & \\
\multicolumn{7}{c}{{\small $U_\textrm{eff}=5.724$~eV
(scaled $2 \times 2$~\cite{linscott2018role})}} \\ \cline{2-4}
& & &  & & & \\ 

\splitbox{\\ \u \& \\ TDDFT (TDA) }  & \splitbox{\\ 2.80 \\ }  & \splitbox{\\ \bf{2.99} \\  } & \splitbox{\\ \bf{3.24} \\ } & \splitbox{\\3.31 \\ } & \splitbox{\\ \bf{3.49} \\ } & \\   \\  \hline
\splitbox{\\ \u \& \\ \uu (TDA) }  & \splitbox{\\ 1.35 \\ \bf{2.99} }  & \splitbox{\\ \bf{1.41} \\ \bf{3.08}} & \splitbox{\\ \bf{1.80} \\ } & \splitbox{\\ \bf{2.12} \\ } & \splitbox{\\ \bf{2.86} \\ } & \\ 
& & & &  &  \\ \hline

& & &  & & & \\
\multicolumn{7}{c}{{\small $U_\textrm{eff}=3.798$~eV
(averaged $1 \times 1$~\cite{linscott2018role})}} \\ \cline{2-4}
& & &  & & & \\ 
\splitbox{\\ \u \& \\ TDDFT (TDA) }  & \splitbox{\\ \bf{2.16} \\ \bf{3.06}}  & \splitbox{\\ \bf{2.45} \\ } & \splitbox{\\ \bf{2.71} \\ } & \splitbox{\\ 2.85 \\ } & \splitbox{\\ \bf{2.98} \\ } & \\   \\  \hline
\splitbox{\\ \u \& \\ \uu (TDA) }  & \splitbox{\\ 0.20 \\ \bf{2.64}}  & \splitbox{\\ 0.35 \\ \bf{2.85}} & \splitbox{\\ \bf{0.56} \\ \bf{2.93}} & \splitbox{\\ \bf{0.90} \\ \bf{3.00}} & \splitbox{\\ \bf{1.02} \\ } & \\ 
& & & &  &  \\ \hline
& & & &  &  \\
Exp. (solvent)~\cite{doi:10.1021/ic400980b}  & \bf{1.10} &  \bf{2.00}  & 
\bf{ 2.15 }& \bf{ 2.45 } & \\   \\  \hline

& & & & &  \\
CASSCF~\cite{doi:10.1021/ic400980b} &  \splitbox{\hlp{0.35} \\ \hlb{1.06} \\ }  & \splitbox{\hlp{0.43} \\ \hlv{2.76} \\ }& \splitbox{\hlp{0.56} \\ \hlv{2.80} \\} & \splitbox{\hlb{0.72} \\ \hlv{2.84} \\}  & \splitbox{\hlb{0.87} \\  \\ }  \\   \hline 
& & & & &  \\
\splitbox{CASSCF \\+NEWPT2~\cite{doi:10.1021/ic400980b}\\} & \splitbox{\hlp{0.49} \\ \hlb{1.36} \\}  & \splitbox{\hlp{0.58} \\ \hlv{2.67} \\} & \splitbox{\hlp{0.75} \\ \hlv{2.72} \\} & \splitbox{\hlb{0.95} \\ \hlv{2.75} \\} & \splitbox{\hlb{1.13} \\  \\} \\ 
\hline \hline
\end{tabular}
}
\end{center}
\caption[Energies (in eV) of the singlet excitations of
CoL$_2$Cl$_2$.]{Energies (in eV) of the singlet excitations of
CoL$_2$Cl$_2$, as obtained  
without symmetry 
assignment from 
the peak positions of Fig.~\ref{fig:co-exc}, with smaller  peaks and shoulders removed.
Coinciding peaks in Fig.~\ref{fig:co-abs}   are assigned as optically bright excitations,  highlighted with a bold font. 
Experimental peak energies extracted from 
Ref.~\citenum{doi:10.1021/ic400980b} are accurate to the nearest  
$0.05$~eV, approximately.
}
\label{tab:c8t8}
\end{table}

\section{Conclusion} 
\label{Section6}

In this work, we carried out a systematic investigation of the extension of
Hubbard $U$ corrected approximate Kohn-Sham DFT to the excited-state
regime, namely TDDFT+$U$.
For this, a linear-scaling, linear-response implementation of TDDFT+$U$
was developed within the \textsc{ONETEP} code,
by combining existing linear-scaling DFT+$U$~\cite{PhysRevB.85.085107,
PhysRevB.73.045110}, 
conduction-band optimization~\cite{PhysRevB.84.165131}, and beyond Tamm-Dancoff
TDDFT~\cite{doi:10.1063/1.4817330,doi:10.1063/1.4936280,doi:10.1021/acs.jctc.5b01014} methods.
Our implementation has allowed us to decouple and 
analyse  the separate and combined effects of Hubbard $U$
correction at the DFT (potential) and TDDFT (kernel) levels, 
offering insights into the performance and potential range
of useful applicability of TDDFT+$U$.
A four-level toy model has proved invaluable 
to our interpretation of TDDFT+$U$ and
the numerical results that support this picture, 
particularly in two representative 
low-spin (spin-unpolarised but non-isoelectronic)  
Ni-centred complexes.
In these systems, we first treated the 
Hubbard $U$ as a free parameter in order 
to understand in detail the exciton binding effect of TDDFT+$U$, as well
as the tendency for the effects of DFT+$U$
and TDDFT+$U$ to approximately cancel. 
We also analysed in detail the differing effects of Hubbard $U$
on TDDFT depending on whether the Tamm-Dancoff approximation 
is invoked.
Including also a challenging Co-centred open-shell, high spin coordination complex,
we calculated first-principles Hubbard $U$ and Hund's $J$
parameters for all three systems, following the spin-polarised, 
minimum-tracking~\cite{moynihan2017self} linear-response approach introduced
in Ref.~\citenum{linscott2018role}.
This has enabled us to generate fully first-principles
excitation and absorption spectra for each of these
elusive systems and to compare with prior experimental and
quantum chemical findings.

Physically, our analysis shows that TDDFT+$U$ can be thought of as a
self-interaction correction for excitons, acting to enhance the exciton
binding.
Indeed, quite apart from \uu being mandated in principle when TDDFT is
 applied upon a \u Kohn-Sham eigensystem, we find that \uu can be very
 effective in re-binding well-defined strongly-localized, optically dark ligand-field
 excitations.
 The Hubbard $U$ dependence of this re-binding is illustrated nicely, we
 think, in Fig.~\ref{fig:nn-uuu-exc}.
  Our study has identified examples of such ligand-field excitations that are 
  predicted at low energies by quantum-chemistry methods but
 pushed to unrealistically high energies by first-principles DFT+$U$.
TDDFT+$U$ can address this effectively, to some extent, but
 only if the localized character
 of those excitations has not already been eradicated by DFT+$U$, however, 
 as illustrated in Fig.~\ref{fig:co-exc}.
In general, while \u shifts excitation energies of transitions into,  out of, 
and within the targeted localised subspace by modifying 
the underlying Kohn-Sham energy levels in 
proportion to the effective Hubbard $U$ parameter,
approximately speaking, 
 \uu only directly affects  transitions \emph{within} that  subspace.
 This gives rise to an incomplete cancellation of the effects of 
 DFT+$U$ and TDDFT+$U$ and 
as a result, we conclude that while the combination of \u and \uu may 
 often give rise to something of a linear combination of the two method's 
 effect, the interaction between them may also be  
 non-trivial, with multiple $U$-dependence regimes potentially emerging.

Existing approaches for the calculating the adiabatic limit of 
the Hubbard $U$ and Hund's $J$ within DFT (or more precisely 
generalised Kohn-Sham DFT, in practice), such as linear-response 
method, already calculate the necessary
parameters for TDDFT+$U$ by construction.
Indeed, our results suggest that these parameters may be \emph{more} suited
to TDDFT+$U$ than to DFT+$U$, in the sense that $U$
(and $J$) exist at the same energy-derivative order as the kernel correction 
$f_U^{\sigma \sigma'}$, whereas the DFT+$U$ correction to the potential
retains a somewhat arbitrary constant (in the sense that a choice
of double-counting correction must be made).
 Furthermore, our results add to the growing body of literature that suggests that 
 DFT+$U$ should be used with caution on closed-shell, or more generally
 low-spin systems, as discussed in Ref.~\citenum{linscott2018role}
 and references therein.
Our findings on the closed-shell complex Ni(CO)$_4$, for example, 
where DFT \& TDDFT+$U$ performs rather well when judged against
the INDO/S quantum chemistry method 
(see third panel of Fig.~\ref{fig:no-abs-u}), suggest a basic failure of the
DFT+$U$ corrective potential in combination with the 
first-principles $U_\textrm{eff} = U - J$.

An interesting avenue for future investigation  in 
problematic systems such as those studies is the use of a 
second Hubbard $U$ parameter to enforce Koopmans' 
condition to the targeted subspace~\cite{PhysRevB.94.220104}, 
as discussed in Section~\ref{mycaveat}.
This idea effectively fixes the arbitrary constant in DFT+$U$, 
or locates the double-counting correction from first principles, 
but its effect in non-trivial systems is yet to be investigated.
Overall, notwithstanding, a picture emerges in the present work whereby
the application of Hubbard $U$ correction to a single localized subspace
alone (with first-principles parameters~\cite{linscott2018role}) 
may be advantageous 
and expedient for the qualitative description of 
optically dark $3d \rightarrow 3d$ excitations that are difficult to otherwise
recover.
This description can come, however, at the expense of considerably worsening
the description of less localized excitations that are well described by standard,
semi-local approximations to TDDFT.
Further research is warranted, therefore, on generalizations to the 
contemporary DFT+$U$ functional such as to incorporate
further chemical information.
More basically, perhaps, 
but no less interestingly, more research is needed on the 
effects of DFT+$U$,
DFT+$U$+$J$~\cite{PhysRevB.84.115108}, 
DFT+$U$+$V$~\cite{matteoUV} (and 
their potential respective 
TDDFT+$U$ extensions) to more
delocalised subspaces centred on ligand atoms 
(see for example the oxygen 
$2p$ treatment in Ref.~\citenum{linscott2018role})
or even bond-centred ones.

\section{Acknowledgements}

We gratefully acknowledge the support of 
Trinity College Dublin's Studentship Award
and  School of Physics.
The authors also acknowledge the DJEI/DES/SFI/HEA Irish Centre for High-End Computing (ICHEC) for the provision of computational facilities and support.
We also acknowledge Trinity Centre for High Performance Computing
(Trinity Research IT)
and Science Foundation Ireland, for the maintenance and
funding, respectively, of the Lonsdale and Boyle clusters on which
further calculations were performed.

\appendix

\section{Computational details}
First-principles simulations were performed using 
our implementation of the TDDFT+$U$
method in the ONETEP linear-scaling package~\cite{doi:10.1063/1.1839852,PSSB:PSSB200541457,PhysRevB.85.085107}.
All calculations used the Perdew-Burke-Ernzerhof (PBE)~\cite{PhysRevLett.77.3865} 
generalized gradient approximation as the underlying
exchange-correlation functional.
Norm-conserving scalar-relativistic PBE pseudo-potentials were generated 
in-house for neutral Ni, Cl, O, C, N, H, 
and Co$^{2+}$ using the OPIUM code~\cite{opium}. 
Ground-state simulations are referred to here 
as \textit{single-point} (SP), and the subsequent 
procedure of variationally optimising the second set of NGWFs to represent the unoccupied manifold~\cite{PhysRevB.84.165131} is referred as \textit{conduction} (COND). 
Initial ionic geometries were adopted from a prior first-principles study~\cite{Demuynck1971522} in the case of Ni(CN)$_4^{2-}$, and from  experimental data~\cite{doi:10.1063/1.437911} in the case of Ni(CO)$_4$.
These molecular geometries were optimized iteratively 
until they fulfilled three convergence criteria: on the maximum atomic displacements ($0.005~a_0$), total energy per atom
($10^{-6}$~Ha), and total atomic force ($0.002$~Ha/$a_0$), by means of the Broyden-Fletcher-Goldfarb-Shanno (BFGS) algorithm~\cite{PhysRevB.83.195102,doi:10.1063/1.4728026}. 
In the case of the  CoL$_2$Cl$_2$,  the molecular geometry was directly adopted from Ref.~\citenum{antti1972metal} for the sake of preserving with comparability of the spectra of Ref.~\citenum{doi:10.1021/ic400980b}, which use the same geometry.
The  molecules were then positioned into smaller cuboidal simulation boxes centred on their respective metallic atoms,  with the available minimum dimensions needed to satisfy the requirements of the Martyna-Tuckeman periodic boundary correction (PBC), which was  applied with its dimensionless parameter set to $7$ as recommended in Ref.~\citenum{doi:10.1063/1.477923}.

\begin{table}[H]
\renewcommand{\arraystretch}{1.4} \setlength{\tabcolsep}{8pt}
\begin{center}
{
\begin{tabular}{lcc}
\hline \hline
Parameter & Stage &  Value  \\ \hline
E\td{\mathrm{cut}} &  All & 1200~eV  \\
R\tdu{NGWF} &  All & 12~a\td{0}   \\
N\tdu{NGWF}{Ni} &  SP (COND) & 9 (18)   \\
N\tdu{NGWF}{Co} &  SP (COND) & 9 (18)   \\
N\tdu{NGWF}{Cl} &  SP (COND) & 4 (13)   \\
N\tdu{NGWF}{C} &  SP (COND) & 4 (8)   \\
N\tdu{NGWF}{N} &  SP (COND) & 4 (8)   \\
N\tdu{NGWF}{O} &  SP (COND) & 4 (8)   \\
N\tdu{NGWF}{H} &  SP (COND) & 1 (2)   \\
 \hline \hline
\end{tabular}}
\end{center}
\caption{The  converged run-time parameters 
used  for  Ni(CN)$_4^{2-}$, Ni(CO)$_4$, and CoL$_2$Cl$_2$.
Here, E\td{\mathrm{cut}} is the kinetic energy cut-off,
R\td{NGWF} is the atom-centred 
nonorthogonal generalized Wannier function (NGWF) spherical cut-off radius,
and N\tdu{NGWF}{} is the number of NGWFs per atom
to be variationally optimized in situ.}
\label{tab:c8t2}
\end{table}

A series  of convergence tests were performed  to 
safeguard the quality excited-state simulations, while maintaining a reasonable computational cost at the SP, COND and TDDFT levels
(recalling that the effective $U$ is treated as a parameter, which significantly 
multiplies the total computational demand of the study). 
The  resulting common set of parameters used in this study is summarized in Table~\ref{tab:c8t2}.
The effective plane-wave kinetic energy  cut-off (E\td{\mathrm{cut}}) and the cut-off  radius (R\td{NGWFs}) of the variationally-optimized 
nonorthogonal generalized Wannier functions
NGWFs, a minimal basis generated by \textsc{ONETEP}, 
were converged at values of $1200$~eV and $12~a_0$, respectively, 
yielding a energy error per atom within $1$~meV in  SP calculations.   
The value of R\td{NGWF}  was separately tested in COND calculations
and found to be adequate for describing the virtual orbital eigen-energies. %
A total of  $9$($18$) spin-degenerate NGWFs 
were used for Ni atoms in order to complete the period up to Kr, and  a total of $4$ NGWFs for each of C, O and N were used to complete the period up to Ar, were optimized at the SP (COND) level in our Ni-centered complexes, whereas 
for the Co-centred complex $9$ ($18$), $4$ ($13$), $4$ ($8$), and  $1$ ($2$) NGWFs  were variationally optimized for Co, Cl, (C,N), and H atoms during SP (COND) simulations
As CoL$_2$Cl$_2$ is an open-shell system, spin-polarized calculations were performed with a fixed total spin of 3 $\mu_\mathrm{B}$,
and the initial configuration of Co for the pseudo-atomic solver 
(which effects both the NGWF initial guess and the $3d$ pseudo-orbitals defining the DFT+$U$
subspace) was set to the theoretical high-spin 
configuration of [Ar]$4s^03d^7$,  with a 3 $\mu_\mathrm{B}$ total spin.
The occupied-unoccupied Kohn-Sham eigenvector product spaces were  constructed by using full valence manifolds, which are represented by $24$ and $25$ spin-degenerate NGWFs in Ni(CN)$_4^{2-}$ and Ni(CO)$_4$, respectively, and $49$ and $46$ NGWFs for spin-up and spin-down, respectively, in  CoL$_2$Cl$_2$. 
For the conduction manifolds, 20 (10 per spin channel), 16 (8 per spin channel) and 11 
(4 for up and 7 for down) KS conduction orbitals were optimized in  Ni(CN)\td{4}\tu{2-}, Ni(CO)\td{4}, and CoL$_2$Cl$_2$, respectively. 
These  parameters  were selected on the basis of KS eigenvalues, providing
sufficiently many bound states for the targeted spectral range in TDDFT calculations.
The first $50$ singlet excitations for Ni-centered complexes and $20$  singlet excitations for CoL$_2$Cl$_2$  were calculated by variational minimization,
within the larger valence-conduction product space spanned by the 
optimized NGWF basis .
We do not place a strong emphasis on the higher-energy 
excitations shown in our plots, being  more interested and 
confident in the lower-energy excitations affected
by the Hubbard $U$ correction.
In particular, in many of our figures the EES and ABS appear gapped at
high energy, but this is nothing more than 
an artefact of the limited number
of excitations calculated.

\section{First-principles calculation of Hubbard \textit{U} and \textit{J} 
parameters using the minimum-tracking linear-response
method}\label{sec:c8s4}

The efficiency and robustness of the DFT+$U$(+$J$) method is essentially dependent on the determination of the Hubbard parameters.
A common approach is to use  linear-response to determine them~\cite{PhysRevB.58.1201,PhysRevB.71.035105}.
In this work, we employ the recently-introduced minimum-tracking variant of linear-response as implemented in the \textsc{ONETEP} code~\cite{moynihan2017self}, and in particular, its spin-polarized extension introduced in Ref.~\citenum{linscott2018role}. 
In this, the `scaled $2\times 2$' method  can be used to 
evaluate the Hubbard $U$, Hund's $J$, and effective Hubbard 
$U$ parameter ($U_\mathrm{eff}=U-J$) for all three systems using the formulae
\begin{align}
U &{}=\frac{1}{2}\frac{\lambda_U \left(f^{\uparrow\uparrow}+f^{\uparrow\downarrow}\right)+f^{\downarrow\uparrow}+f^{\downarrow\downarrow}}{\lambda_U+1} \label{eq:c4e16}  \\
\mbox{and} \quad J &{}=-\frac{1}{2}\frac{\lambda_J \left(f^{\uparrow\uparrow}-f^{\downarrow\uparrow}\right)+f^{\uparrow\downarrow}-f^{\downarrow\downarrow}}{\lambda_J-1},  \label{eq:c4e17}  
\end{align}
where 
\begin{align}
\lambda_U=\frac{\chi^{\uparrow\uparrow}+\chi^{\uparrow\downarrow}}{\chi^{\downarrow\uparrow}+\chi^{\downarrow\downarrow}},  
\;\;\; \mathrm{and}\;\;\;
\lambda_J=\frac{\chi^{\uparrow\uparrow}-\chi^{\uparrow\downarrow}}{\chi^{\downarrow\uparrow}-\chi^{\downarrow\downarrow}}.
\end{align}
The spin-dependent
interaction strengths $f^{\sigma \sigma'}$
are calculated by incrementally varying subspace-uniform
perturbatimg potentials $\delta v^{\sigma}_\mathrm{ext}$, 
relaxing fully to the ground-state on each step, and then
measuring the resulting small changes 
in the subspace occupancies 
$n^\sigma$ and subspace-averaged Kohn-Sham 
potentials $v_{\mathrm{KS}}^{\sigma}$.
The projected interacting response matrices are given by
$\chi^{\sigma \sigma'} = d n^\sigma / d v^{\sigma'}_\mathrm{ext}$.
When the interaction strengths $f^{\sigma \sigma'}$
are calculated using a $2 \times 2$ matrix equation
indexed by spin, we arrive at the `scaled $2 \times 2$' model,
which reproduces conventional formulae for $U$ and $J$.
Indeed, for spin-unpolarized systems such as the Ni-centered complexes 
studied in this work, $\lambda_U=1$ and $\lambda_J=-1$, 
and as a result we have $U = \left( f^{\sigma \bar{\sigma}} + f^{\sigma \sigma}\right) / 2 $, 
$J = \left( f^{\sigma \bar{\sigma}} - f^{\sigma \sigma}\right) / 2 $, 
and, simply but reassuringly, $U_\textrm{eff} = f^{\sigma \sigma}$.

When spin-off-diagonal elements are neglected, instead, 
we have the `averaged $1 \times 1$' model, in which  
$U_\mathrm{eff}=(U^\uparrow+U^\downarrow)/2$, 
where $U^\sigma =
d \left(  v^{\sigma}_\mathrm{KS} - 
v^{\sigma}_\mathrm{ext} \right) / d n^\sigma$.
This model effectively decouples the spin 
populations into distinct sites, reflecting the 
form of the canonical \u functional.
Each spin channel, for a given localized subspace, then forms part of the screening bath for the  other, and 
the effects of Hund's $J$ are then already 
incorporated into $U_\mathrm{eff}$ at an approximate level.

In practice, a discrete logarithmic grid of
perturbation strengths, 
$\{-0.10, \;-0.01,\; 0.00, \;0.01,\; 0.10\}$~eV, 
was used in this work 
to calculate the  $U$ and $J$ parameters, resulting
in excellent linear fits.
For the spin-unpolarized Ni-centred complexes, it was
necessary only to perturb one spin channel, since half
of the spin-indexed matrix elements could be filled
using symmetry.
The resulting parameters  are summarized in Table~\ref{tab:c8t3}. 
\begin{table}[H]
\renewcommand{\arraystretch}{2.0} \setlength{\tabcolsep}{10pt}
\begin{center}
{
\begin{tabular}{lll}
\hline \hline 
 Interaction  &   Ni(CN)$_4^{2-}$& Ni(CO)$_4$ \\ \hline

$f^{\sigma \sigma}$,  $f^{\sigma \bar{\sigma}}$ &   6.901, 8.456   &  9.849, 11.388   \\

 $U$,  $J$  &  7.678, 0.777    &    10.618, 0.769  \\
 
 $U_{\mathrm{\mathrm{eff}}}$  &  \bf{6.901}   &  \bf{9.849}    \\

 \hline \hline
\end{tabular}}
\end{center}
\caption{Hubbard and Hund parameters (in eV) calculated using the scaled $2\times 2$ method of  Ref.~\citenum{linscott2018role}.}
\label{tab:c8t3}
\end{table}

As  CoL$_2$Cl$_2$ is a spin-polarized system, the responses of each spin channel were measured by perturbing the respective spin channels, separately, one at a time.
The resulting first-principles parameters for the Co $3d$ subspace are summarised in Table~\ref{tab:c8t4}. 

\begin{table}[H]
\renewcommand{\arraystretch}{2.0} \setlength{\tabcolsep}{30pt}
\begin{center}
{
\begin{tabular}{ll}
\hline \hline 
 Interaction  &  CoL$_2$Cl$_2$ \\ \hline

$f^{\uparrow\uparrow}$,  $f^{\downarrow\downarrow}$ &  13.711, 15.268 \\

$f^{\downarrow\uparrow}$,  $f^{\uparrow\downarrow}$ &   7.650, 6.029 \\

$\lambda_U$, $\lambda_J $ & -0.039, -0.195 \\

 $U$, $J$  &  6.529, 0.805     \\

 $U_{\mathrm{\mathrm{eff}}}$  & \bf{5.724}    \\ 
\hline
  $U^\uparrow$,  $U^\downarrow$  & 3.503, 4.093  \\
  
 $U_\mathrm{eff}$  &  \bf{3.798}   \\

 \hline \hline
\end{tabular}}
\end{center}
\caption{Hubbard and Hund parameters (in eV) 
calculated using the scaled $2\times 2$ (top panel) and the averaged $1\times 1$ 
methods 
(bottom panel) of Ref.~\citenum{linscott2018role} for  CoL$_2$Cl$_2$.}
\label{tab:c8t4}
\end{table}

\end{document}